\documentclass[a4paper,11pt]{article}
\pdfoutput=1 
\usepackage{mathtools}
\usepackage{tikz-cd}
\usepackage{jheppub} 

\usepackage[T1]{fontenc} 
\usepackage{dsfont}
\usepackage{tensor}
\usepackage{setspace}
\usepackage[style=numeric,sorting=none]{biblatex} 
\title{\boldmath Two Dimensional Conformal Field Theory and a Primer to Chiral Algebras}

\author[1,2]{Joaquin Liniado}

\affiliation[1]{Department of Applied Mathematics and Theoretical Physics \\
Cambridge University, CB3 0WA, UK}
\affiliation[2]{Instituto de Física La Plata\\
CONICET and Universidad Nacional de La Plata\\
CC 67 La Plata, Argentina}

\emailAdd{jl2185@cantab.ac.uk}

\abstract{We review various aspects of two dimensional conformal field theories paying close attention to the algebraic structures that intervene. We provide a compact description regarding the appearance of a chiral algebra as the symmetry algebra related to local conformal symmetry, namely the Virasoro algebra. We then introduce two dimensional conformal field theories with additional symmetries in which extended chiral algebras emerge as a natural generalization of the conformal case.}

\begin{document} 
\maketitle
\flushbottom

\section{Introduction}

Over the past few decades, the study of two dimensional systems with conformal symmetry has enjoyed a privileged status in the theoretical physicist's agenda. Indeed, conformal invariance in two dimensions plays a central role in the description of second order phase transitions in statistical models \cite{Polyakov2, Wilson, Zuber, Itzikson, Cardy3, Belavin} as well as in the  understanding of fundamental aspects of string theory \cite{Polyakov4,Polyakov5, Belavin, Neveu, Ramond, Friedan}. From a slightly more mathematical perspective, it has also inspired exceptional contributions in the theory of vertex operator algebras and $\mathcal{W}-$algebras \cite{Moore, Zamolodchikov1, Goddard2, Verlinde, Drinfeld, Bais1, Bais2}.


All of these examples have two important features in common: Conformal symmetry in two dimensions, and the fact that something else (different in each case) is required to fully describe the problem. This implies that studying conformal symmetry \textit{alone} is not of much real interest, but at the same time, it makes the choice of what to discuss and what not to, a surprisingly complex task. In particular, for the topic we are hereby concerned, namely \textit{chiral algebras}, the literature appears to be dissected. Either conformal invariance is treated in detail and little reference is made to chiral algebras in general, or else, the mathematics of conformal symmetry is taken to be understood and the study of chiral algebras begins under that assumption. A few notable exceptions should be made, and that is the case of the standard texs \cite{Francesco, Blumenhagen, Ginsparg} in which, loyal to their own style, both topics are treated thoroughly. Nonetheless, there is so much more covered in one and the other, that the transition from conformal symmetry to chiral algebras may not be apparent. 

It is the aim of this review article to provide a mathematically rigorous introduction to two dimensional conformal field theories, in which the study of chiral algebras appear as a natural generalization. As such, we present the usual standard material regarding conformal symmetry \cite{Francesco,Ginsparg, tong, Polchinski, GSWitten, Alvarez, Cardy, Friedan2, Blumenhagen, Zwiebach2,Ribault} with two notable differences: In the first place, we cover some mathematical aspects that are usually avoided in the physics literature, and we do so by addressing specific objects which may give rise to confusion. Second and most importantly, we structure our presentation in a \textit{story-like} fashion in which all of the elements are consistently related to each other. Our focus is centered on the abstract algebraic construction which enables the full fledged capitalization of the infinite dimensionality of the symmetry algebra. In other words, we carefully describe the mechanism in which the infinite dimensional symmetry can be realized in the quantum theory through the Virasoro algebra, and outline its extensive consequences. This perspective gives rise to the study of chiral algebras as the natural generalization of the aforementioned construction. In this way, this article may serve as a self contained introduction to two dimensional conformal field theory, but also, as a primer in the study of chiral algebras.  

Having argued why we consider this review may come in handy for the reader who is interested in two dimensional conformal field theories and chiral algebras, let us devote some time to advertise the topic. To begin with, it is worth noting that the term \textit{chiral algebra} is used in the literature with slightly different meanings depending on the author. Throughout the text we will try to make its definition as precise as possible, but heuristically speaking, we will think of a chiral algebra as the symmetry algebra of a two dimensional system with (at least) conformal symmetry. If the system contains additional symmetries other than conformal symmetry, we will still refer to the symmetry algebra of the system as a chiral algebra, but in this case, it will be enlarged. 

From a mathematical point of view, the main reason why conformal symmetry in two dimensions is so powerful, is the fact that the corresponding symmetry algebra is infinite dimensional. The existence of a symmetry in a quantum field theory gives rise to the so known Ward Identities which can be informally understood as the quantum version of Noether's theorem. The infinite dimensionality of the symmetry algebra implies an infinite number of constraints on the system captured through the Ward identities, which considerably simplifies the analysis, providing in some cases the possibility to exactly solve a theory with symmetry arguments only. A remarkable aspect of two dimensional CFT's is the possibility of having extended symmetries other than conformal symmetry itself. This can be motivated from a physical perspective, given that for certain applications in string theory or statistical mechanics an additional symmetry is required for a good description of the system, but at the same time, extended symmetries can be used to facilitate the analysis of a large class of conformal field theories (called rational conformal field theories) and, eventually, to classify certain types of conformal field theories.

In this article we will cover all of the preliminary material concerning conformal symmetry in two dimensions, paying close attention to the algebraic structures that intervene. With the intention of relating in a chronological order all of the agents involved in the formalism, we focus on the conceptual construction, giving less importance to specific examples. This will allow us to understand the nuances behind the appearance of the Virasoro algebra as the symmetry algebra of a two dimensional conformal field theory, and its far reaching implications. In particular, we put special emphasis on the mechanism used to realize the symmetry in the quantum theory, by structuring our Hilbert space in terms of irreducible representations of the symmetry algebra. This perspective will (hopefully) provide enough insight to stimulate the use of extended symmetries  as a device to further constraint a system, which leads to the systematic study of chiral algebras.

The organization of this paper is as follows. In \S \ref{sec:conformalsymmetry} we discuss conformal symmetry in both $d=2$ and $d\geq 3$ dimensions. We do so in an unusual format: We first introduce the topic following the standard presentation, and we then complement each section (corresponding to $d=2$ and $d\geq 3$) with some extra mathematics in order to address some of the aspects that may give rise to confusion. The reason we proceed in this way is to make the material as accesible as possible. In \S \ref{sec:3dCFT} we discuss conformal field theories in $d \geq 3$ dimensions, restricting ourselves to the ideas that can be applied to the two dimensional case as well. This will highlight the contrast between the former and the latter. In \S \ref{sec:2dCFT} we initiate our study of two dimensional conformal field theories and we do so in detail. We start analyzing the transformation properties of correlation functions under conformal symmetry, which leads naturally to the introduction of the \textit{conformal Ward identities}. We then introduce a powerful tool to conveniently express the symmetry information contained in the Ward identities known as the \textit{operator product expansion} (OPE). We proceed by introducing a quantization prescription known as \textit{radial quantization} which enables the transition from a field theory description to an operator formalism. In this context, we translate the symmetry statements regarding the transformation of correlation functions into commutator language, which leads to the realization of the symmetry algebra at the quantum level. At this point we decompose our Hilbert space in terms of irreducible representations of the symmetry algebra and we investigate the consequences, stressing the power of the constraints imposed by conformal symmetry only. This purely algebraic perspective leads naturally to the introduction of systems with extended symmetries in \S \ref{sec:chiralalgebra} in which the notion of general chiral algebras is presented.


\section{Conformal Symmetry}
\label{sec:conformalsymmetry}

Conformal symmetry is a mathematically subtle topic if we want to introduce it in a rigorous manner . In particular, given that the interest of studying conformal transformations has it's origins in the realm of physics \cite{Bateman, Cunningham}, most of the literature was written by physicists, which as we very well know, are not particularly interested in the mathematical rigor expected by mathematicians. The main reason why this is so, is because it can very well happen that mathematical precision may appear more as a burden than as a useful tool. Conformal symmetry, in this perspective, is an exotic example: Mathematical rigor does not provide a revealing insight, but the lack of it, can lead to confusion at the conceptual level. 

Of course, an accurate mathematical description always comes with a price. On the one hand, a use of language that we physicists are not acquainted with. On the other hand, a seemingly unnecessary amount of lines to define something that could be simply expressed in a couple of characters. In this article, we choose to pay this price but not without a strategy. We will try to address the issues that may cause dissatisfaction when introduced from the physicist's point of view, and those issues only. In order for the presentation to be as friendly as possible, we will proceed as follows. We introduce the topics following standard literature \cite{Francesco, Ginsparg}, pointing out specific statements or definitions that can be more accurately presented. We then include an additional section in which this ideas are discussed with more care following \cite{Schottenloher}. The reader who is satisfied with the first of the options, may choose to avoid the technicalities and jump directly to the following section. In either case, the treatment will eventually converge to the usual physics dictionary. 

\subsection{Conformal Symmetry in $d \geq 3$ Dimensions}
\label{sec:conformaltransformations}
Let $(M,g)$ be a $d$-dimensional semi Riemannian manifold with $d\geq 3$\footnote{The $d=2$ case will be studied separately in the next section.}, with signature $(p,q)$. Under a coordinate transformation $x \mapsto x'$, we have 
\begin{equation}
\label{ec:metrictransformation}
    g_{\mu \nu}\mapsto g'_{\mu \nu}=\frac{\partial x^\rho}{\partial x'^{\mu}}\frac{\partial x^{\sigma}}{\partial x'^{\nu}}g_{\rho \sigma}\,.
\end{equation}
We say that the transformation $x \mapsto x'$ is conformal
if the metric remains invariant up to a scale factor: 
\begin{equation}
\label{ec:conformalcondition}
    g_{\mu \nu}'=\Omega^2(x)g_{\mu \nu}\,,
\end{equation}
where $\Omega:M \to \mathbb{R}_+$ is a smooth function. To study the structure of conformal transformations, it is customary to start by considering infinitesimal transformations and only then analyzing the general (finite) case. To simplify our discussion we take our manifold to be $M=\mathbb{R}^{p,q}$, although everything can be equally  done in a general semi-Riemannian manifold. Under an infinitesimal transformation $x^\mu \mapsto x'^\mu=x^\mu + \epsilon^\mu(x)$, equation \eqref{ec:metrictransformation} becomes, to first order in $\epsilon$:
\begin{equation}
    g'_{\mu \nu}=g_{\mu \nu}-(\partial_\mu \epsilon_\nu +\partial_\nu \epsilon_\mu)\,.
\end{equation}
The conformal condition given in equation \eqref{ec:conformalcondition} implies that:
\begin{equation}
\label{ec:francescokilling}
    \partial_\mu \epsilon_\nu +\partial_\nu \epsilon_\mu = f(x)g_{\mu \nu}
\end{equation}
for some $f:\mathbb{R}^{p,q}\to \mathbb{R}$. We now make a further simplification and take $g_{\mu \nu}=\eta_{\mu \nu}$ to be the flat metric. To determine $f(x)$ we start by tracing both sides:
\begin{equation}
\label{ec:conformal4}
    f(x)=\frac{2}{d}\partial_\mu \epsilon^\mu \,,
\end{equation}
and we proceed by acting with an extra derivative on equation \eqref{ec:francescokilling}, permuting the indices, and taking a linear combination to get:
\begin{equation}
\label{ec:conformal3}
    2\partial_\mu \partial_\nu \epsilon_\rho = \eta_{\mu \rho}\partial_\nu f+\eta_{\nu \rho}\partial_\mu f -\eta_{\mu \nu} \partial_\rho f\,,
\end{equation}
which upon contracting with $\eta^{\mu \nu}$ we obtain:
\begin{equation}
\label{ec:killing2d}
    2\partial^2 \epsilon_\mu = (2-d)\partial_\mu f \,.
\end{equation}
Next, we act with $\partial_\nu$ on this expression and $\partial^2$ on equation \eqref{ec:francescokilling} to get:
\begin{equation}
\label{ec:conformal1}
    (2-d)\partial_\mu \partial_\nu f=\eta_{\mu \nu}\partial^2 f
\end{equation}
which upon contraction with $\eta^{\mu \nu}$ once again, we find:
\begin{equation}
\label{ec:conformal2}
    (d-1)\partial^2 f = 0\,.
\end{equation}
Since we are considering the $d\geq 3$ case, we note that equations \eqref{ec:conformal1} and \eqref{ec:conformal2} imply that $\partial_\mu \partial_\nu f = 0$. Hence, we have that:
\begin{equation}
\label{ec:solucionf}
    f(x)=A + B_\mu x^\mu\,.
\end{equation}
If we substitute this expression for $f$ in equation \eqref{ec:conformal3} we see that $\partial_\mu \partial_\nu \epsilon_\rho$ is constant, which implies that $\epsilon_\mu$ is at most quadratic in the coordinates:
\begin{equation}
    \epsilon_\mu =a_\mu + b_{\mu \nu}x^\nu + c_{\mu \nu \rho}x^\nu x^\rho
\end{equation}
with $c_{\mu \nu \rho}$ symmetric in its last two indices. Now, since the constraints for $\epsilon_\mu$ given in equations \eqref{ec:francescokilling}-\eqref{ec:conformal3} hold for every $x$, we may treat each term separately. The first term $a_\mu$ is in fact free of constraints and corresponds to an infinitesimal translation. Replacing the linear term in \eqref{ec:francescokilling} we get:
\begin{equation}
    b_{\mu \nu} + b_{\nu \mu} = \frac{2}{d}b^{\rho}{}_\rho \eta_{\mu \nu}\,,
\end{equation}
which implies that $b_{\mu \nu}$ is the sum of an antisymmetric part and a pure trace part:
\begin{equation}
    b_{\mu \nu}= \alpha \eta_{\mu \nu} + m_{\mu \nu} \quad \text{with} \quad m_{\mu \nu}=-m_{\nu \mu}\,.
\end{equation}
The pure trace part corresponds to an infinitesimal scaling transformation, while the antisymmetric part, to an infinitesimal rotation. Finally, if we substitute the quadratic term into equation \eqref{ec:conformal3}, we get:
\begin{equation}
    c_{\mu \nu \rho}= \eta_{\mu \rho}b_\nu + \eta_{\mu \nu} b_\rho - \eta_{\nu \rho}b_\mu\,, \quad \text{with} \quad b_\mu = \frac{1}{d}c^{\lambda}{}_{\lambda \mu}\,.
\end{equation}
The infinitesimal transfomartion corresponding to this term is thus,
\begin{equation}
    x'^\mu=x^\mu + 2(x\cdot b)x^\mu -b^\mu x^2\,,
\end{equation}
which corresponds to an infinitesimal \textit{special conformal transformation} (SCT). The intuitive nature of these transformations is less clear\footnote{The finite version of this transformation is in this regard, somehow more appealing.} and not important for the current analysis. The finite versions of these transformations can be obtained by exponentiation\footnote{See \S \ref{sec:mathperspective1} for a discussion regarding the finite version of an infinitesimal transformation.} and are the following:
\begin{enumerate}
    \item A translation: $x^\mu \mapsto x^\mu + a^\mu$ with $a^\mu \in \mathbb{R}^{p,q}$
    \item A Lorentz transformation: $x^\mu \mapsto M^\mu{}_\nu x^\nu$ with $M^\mu{}_\nu \in SO(p,q)$
    \item A dilatation: $x^\mu \mapsto \lambda x^\mu$ with $\lambda \in \mathbb{R}$ 
    \item A SCT: $x^\mu \mapsto \displaystyle \frac{x^\mu -  b^\mu x^2}{1 - 2b\cdot x + x^2 b^2}$\,.
\end{enumerate}
In particular, we find that an arbitrary finite SCT is the composition of an inversion $x\mapsto 1/x$ followed by a translation $x \mapsto x+b$ followed by another inversion. The conformal factor in equation \eqref{ec:conformalcondition} for the translation and the Lorentz transformation is $\Omega^2(x) = 1$, for the dilatation is $\Omega^2(x)=\lambda^d$ and for the special conformal transformation $\Omega^2(x)=(1-2b\cdot x+b^2x^2)^2$. The set of conformal transformations forms a group\footnote{See \S \ref{sec:mathperspective1} for a precise definition of the conformal group in $d\geq 3$} with respect to composition which can be shown to be isomorphic to $SO(p+1,q+1)$. Furthermore, from these transformations, we may read off\footnote{See \S \ref{sec:mathperspective1} for a discussion of the relation between infinitesimal transformations and generators of the associated Lie algebra.} the form of the generators of the conformal algebra, which are given by:
\begin{enumerate}
    \item Translation: $P_\mu = -i\partial_\mu$
    \item Lorentz transformation: $M_{\mu \nu} = i(x_\mu\partial_\nu - x_\mu \partial_\nu)$
    \item Dilatation: $D=-ix^\mu\partial_\mu$
    \item SCT: $K_\mu=-i(2x_\mu x^\nu \partial_\nu-x^2\partial_\mu)$ \,.
\end{enumerate}
These generators obey the following commutation relations:
\begin{equation}
\label{ec:CRconformalalgebra}
    \begin{array}{c}
         [M^{\mu \nu}, M^{\rho \sigma}]=\eta^{\nu \rho}M^{\mu \sigma} + \eta^{\mu \sigma}M^{\nu \rho}-\eta^{\mu \rho}M^{\nu \sigma}-\eta^{\nu \sigma}M^{\mu \rho} \,, \\
        \cr [M^{\mu \nu},P^\sigma]=\eta^{\nu \sigma}P^\mu - \eta^{\mu \sigma}P^\nu \,, \\
        \cr [M^{\mu \nu},K^\sigma]=\eta^{\nu \sigma}K^\mu - \eta^{\mu \sigma}K^\nu \,, \\
         \cr [D,P^\mu]=P^\mu\,, \\
        \cr [D,K^\mu]=-K^\mu\,, \\
        \cr [K^\mu,P^\nu]=2\left(\eta^{\mu \nu}D-M^{\mu \nu}\right)\,, 
    \end{array}
\end{equation}
where unspecified commutators vanish. The dimension of the conformal algebra in $d$ dimensions is given by $\frac{1}{2}(d+1)(d+2)$ which can be seen from:
\begin{equation}
    \underbrace{\text{Conformal Transf}}_{\frac{1}{2}(d+1)(d+2)}=\underbrace{\text{Translations }}_{d}+\underbrace{\text{ Lorentz }}_{\frac{1}{2}d(d-1)}+\underbrace{\text{ Dilatations }}_{1}+\underbrace{\mathrm{ SCT}}_{d}
\end{equation}
In fact, this is precisely the number of free parameters of the algebra $\frak{so}(p+1,q+1)$, to which the conformal algebra is isomorphic. Notably, we observe that for $d\geq 3$ both the conformal group and the conformal algebra are finite dimensional. 

\subsection{Mathematical Aspects in $d\geq 3$}
\label{sec:mathperspective1}

As mentioned in the introduction of this section, the objective of this subsection is to complement some of the elements introduced in the previous section with some extra mathematics. Namely, the exponentiation of infinitesimal conformal transformations, the generators of the conformal algebra and a proper definition of the conformal group. To do so in a precise manner, some technical definitions will be required, so before going into details, we summarize the main ideas. We will show that the concept of infinitesimal transformations may be well understood in terms of vector fields and integral curves. The infinitesimal variation represented by $\epsilon$ in the previous subsection, will now be described by a vector field so that the corresponding finite transformation will be defined in terms of the associated one parameter group of diffeomorphisms. On the other hand, by expressing infinitesimal transformations in terms of vector fields, we automatically have a well defined vector space structure, that equipped with the commutator between vector fields defines a Lie algebra structure. This perspective provides the relation between infinitesimal transformations and the Lie algebra of conformal transformations. Finally, to define the conformal group, we introduce the concept of a \textit{conformal compactification}. For $d \geq 3$ dimensions, this is a mere technicality to obtain a well defined structure with no trascendental consequences. However, in $d=2$ where the distinction between global and local transformations becomes essential, a proper definition of the (global) conformal group can come in handy. 

So let's start by considering a $d$-dimensional semi Riemannian manifold $(M,g)$ with $d\geq 3$ and signature $(p,q)$. Let $U,V \subseteq M$ be open subsets of $M$. A smooth mapping $\phi:U \to V$ is called a conformal transformation if there exist a smooth function $\Omega:U\to \mathbb{R}_+$ such that:
\begin{equation}
\label{ec:schotenconformal}
    \phi^* g = \Omega^2 g \,,
\end{equation}
where $\phi^* g$ is the pull-back of the metric tensor $g$ by $\phi$, defined by the action $\phi^*g(X,Y)=g({\rm d}\phi(X),{\rm d}\phi(Y))$ for any pair of vector fields $X,Y \in TM$ with ${\rm d}\phi$ the differential of $\phi$. In a coordinate basis, we have for any point $p\in U$:
\begin{equation}
    (\phi^*g)_{\mu \nu} (p)= \left(g_{\rho \sigma}(\phi(p))\right)\partial_\mu \phi^\rho \partial_\nu \phi^\sigma \,,
\end{equation}
so that $\phi$ is a conformal transformation if:
\begin{equation}
   \Omega^2g_{\mu \nu} = (g_{\rho \sigma}\circ \phi)\partial_\mu \phi^\rho \partial_\nu \phi^\sigma \,.
\end{equation}
Let us proceed to classify conformal transformations with an infinitesimal argument. Let $\xi:M\to TM$ be a smooth vector field, i.e, for every $p \in M$, $\xi(p)=\xi_p \in T_p M$ is a tangent vector. The local one parameter group $\phi^{\xi}(t,p)$ associated to $\xi$ is a family of maps $\phi^{\xi}:\mathbb{R}\times M\to M$ that satisfy the flow equation:
\begin{equation}
    \frac{d}{dt}\phi^\xi(t,p)|_{t=0}=\xi(p) \,.
\end{equation}
In particular, for every $p\in M$, the curve $\phi^{\xi}(\,\cdot\,,p):\mathbb{R}\to M$ is the integral curve of $\xi$ through the point $p$. Furthermore, for fixed $t$ the map $\phi^\xi(t,\,\cdot\,):M\to M$ is a local diffeomorphism. We say that a vector field $\xi$ is a \textit{conformal Killing vector field} if $\phi^\xi(t,\,\cdot\,)$ is a conformal mapping for every $t$ in a neighbourhood of $0$. The contact with the previous section comes through the following theorem. Let $\xi= \xi^\mu \partial_\mu$ be a conformal Killing vector field, then there exist a smooth function $\kappa:M\to \mathbb{R}$ such that:
\begin{equation}
\label{ec:confvectorfield}
    \partial_\mu \xi_\nu + \partial_\nu \xi_\mu = \kappa g_{\mu \nu}\,.
\end{equation}
The smooth function $\kappa$ is called a \textit{conformal Killing factor}. We note that equation \eqref{ec:confvectorfield} is exactly the same than \eqref{ec:francescokilling} and thus, the infinitesimal variation $\epsilon^\mu$ is precisely the $\mu$-component of a conformal Killing vector field. The classification of conformal transformations follows in a very similar manner; in the same way than before, we take $M=\mathbb{R}^{p,q}$ and $g_{\mu \nu}=\eta_{\mu \nu}$ to be the flat metric. Then, as in equation \eqref{ec:solucionf} we have that:
\begin{equation}
\label{ec:formakappa}
    \kappa(x)= \lambda + 4 b_\mu x^\mu \quad \text{with} \quad \lambda\in \mathbb{R}\,, b \in \mathbb{R}^{p,q}\,,
\end{equation}
where the factor $4$ is purely conventional. To classify conformal transformations, we analyze the various possibilities for $\kappa$ arising from equation \eqref{ec:formakappa}. Let us start with the easiest case, $\kappa=0$. Replacing in equation \eqref{ec:confvectorfield} we get the usual Killing equation, whose solutions we already know: They are the Killing vector fields associated to translations and orthogonal transformations, given by
\begin{equation}
\label{ec:lorentztrans}
    \xi^\mu(x) = c^\mu + \omega^\mu{}_\nu x^\nu
\end{equation}
with $c^\mu \in \mathbb{R}^{p,q}$  and $\omega_{\mu \nu}$ a real valued rank two tensor. Taking $\omega^{\mu}{}_\nu = 0$, we look for the one parameter group of diffeomorphisms $\phi^\xi(x,t)$ associated to the Killing vector field $\xi^\mu(x)=c^\mu$. We recall that this is given by the flow of the differential equation 
\begin{equation}
  \displaystyle\frac{dx^\mu}{dt}=c^\mu
\end{equation}
whose solution is precisely:
\begin{equation}
    \phi^\xi(t,x) = x + t c \,.
\end{equation}
The associated conformal transformation, corresponding to $t=1$, is the translation:
\begin{equation}
    \phi^\xi(x^\mu) = x^\mu + c^\mu \,.
\end{equation}
Similarly, for $c^\mu=0$, replacing $\xi^\mu(x)=\omega^{\mu}{}_\nu x^\nu$ in equation \eqref{ec:confvectorfield} with $\kappa=0$, we have:
\begin{equation}
    g_{\nu\rho} \omega^\rho{}_\mu +g_{\mu \rho } \omega^\rho{}_\nu=0
\end{equation}
from where we conclude that $\omega_{\mu \nu}$ is an antisymmetric rank two tensor, and thus an element of $\frak{o}(p,q)$. The one parameter group of diffeomorphisms associated to this vector field is precisely:
\begin{equation}
    \phi^\xi (t,x) = e^{\omega t}x
\end{equation}
and the associated conformal transformation is given by:
\begin{equation}
    \phi^\xi(x^\mu) = e^{\omega} x = \Lambda x \quad \text{with} \quad \Lambda \in O(p,q) \,. 
\end{equation}
Hence, we find that this transformation corresponds to an orthogonal transformation. 

Next, we consider the case in which $\kappa \neq 0$, and we start taking $\kappa=\lambda \in \mathbb{R}-\{0\}$. The conformal Killing fields satisfying equation \eqref{ec:confvectorfield} with $\kappa = \lambda$ are given by:
\begin{equation}
\label{ec:dilatation}
    \xi^\mu(x) = \lambda x^\mu\,,
\end{equation}
whose associated one parameter group of diffeomorphisms is $\phi^\xi(x,t)=e^{\lambda t} x$. The corresponding conformal transformation is therefore:
\begin{equation}
    \phi^\xi(x^\mu) = e^\lambda x^\mu
\end{equation}
which are dilatations. All of the conformal transformations introduced up to this point correspond to globally defined one parameter groups of diffeomorphisms. This is not true for \textit{special conformal transformations}, corresponding to the case in which $\kappa = 4 b_\mu x^\mu$. For this value of $\kappa$, equation \eqref{ec:confvectorfield} can be solved to obtain:
\begin{equation}
\label{ec:specialct}
    \xi^\mu (x) = 2 x^\nu x^\mu b_\mu - b^\mu x^2 \,.
\end{equation}
Notably, these Killing conformal vector fields do not admit a globally defined one parameter group of diffeomorphisms, but instead, a local one given by:
\begin{equation}
    \phi^\xi(t,x)=\frac{x^\mu - t b^\mu x^2}{1 - 2tb_\mu x^\mu + x^2 t^2 b^2} \quad \text{with} \quad t \in [t^-,t^+]
\end{equation}
where $[t^-,t^+]$ is some interval around $t=0$. The conformal transformation associated to this local one parameter group is given by:
\begin{equation}
    \phi^\xi(x^\mu)= \frac{x^\mu -  b^\mu x^2}{1 - 2b_\mu x^\mu + q^2 b^2} \,,
\end{equation}
and with this transformation, we have exhausted all of the possible values of $\kappa$. 

It is worth emphasizing that we have described two different type of transformations: Inifinitesimal transformations, using conformal Killing vector fields, and finite transformations, using the flow of these vector fields. Let us summarize our results separately.

For conformal Killing vector fields, from equations \eqref{ec:lorentztrans}, \eqref{ec:dilatation} and \eqref{ec:specialct}, we may conclude that any conformal Killing vector field may be expressed as:
\begin{equation}
    \xi^\mu(x)=c^\mu + \omega^\mu{}_\nu x^\nu + \lambda x^\mu +2 x^\nu x^\mu b_\mu - b^\mu x^2 \,.
\end{equation}
We know that the vector space of smooth vector fields defined over $\mathbb{R}^{p,q}$, equipped with the Lie bracket is a Lie algebra. Notably, it can be shown that the Lie bracket between any two conformal Killing vector fields is again a conformal Killing vector field and thus, the set of conformal Killing vector fields, equipped with the Lie bracket forms a Lie subalgebra. In fact, this Lie (sub)algebra, which we call the \textit{conformal Lie algebra} and we denote it by ${\rm conf}(\mathbb{R}^{p,q})$, is isomorphic to $\frak{o}(p+1,q+1)\cong\frak{so}(p+1,q+1)$. The definition of the generators of the conformal algebra is now transparent; 
\begin{enumerate}
    \item Translations: 
    \begin{equation}
        \xi = c^\mu \partial_\mu \quad \Rightarrow \quad P_\mu = - i\partial_\mu
    \end{equation}
    \item Lorentz transformations: 
    \begin{equation}
        \xi = \omega^\mu{}_\nu x^\nu \partial_\mu=\frac{1}{2}\omega^{\mu \nu}(x_\nu \partial_\mu - x_\mu \partial_\nu) \quad \Rightarrow \quad M_{\mu \nu}=i(x_\mu \partial_\nu - x_\nu \partial_\mu) 
    \end{equation}
    \item Dilatations: 
    \begin{equation}
        \xi = \lambda x^\mu \partial_\mu \quad \Rightarrow \quad D=-ix_\mu \partial^\mu 
    \end{equation}
    \item Special conformal transformations: 
    \begin{equation}
        \xi = 2 x^\nu x^\mu b_\mu\partial_\nu - b^\mu x^2 \partial_\mu \quad \Rightarrow \quad  K_\mu =-i(2x_\mu x^\nu \partial_\nu -x^2\partial_\mu) 
    \end{equation}     
\end{enumerate}
where the imaginary prefactors are purely conventional. Of course, these coincide with the ones defined in the previous subsection and thus, they satisfy the same commutation relations. The virtue of this procedure relies on understanding how infinitesimal transformations may be described in a precise manner using some elementary differential geometry.

On the other hand, we have seen that the most general conformal transformation of the coordinates is a composition of:
\begin{enumerate}
    \item A translation: $x \mapsto x + c$ with $c \in \mathbb{R}^{p,q}$
    \item An orthogonal transformation: $x \mapsto \Lambda x$ with $\Lambda \in O(p,q)$
    \item A dilatation: $x \mapsto \lambda x$ with $\lambda \in \mathbb{R}$
    \item A special conformal transformation (SCT): $x \mapsto \displaystyle \frac{x^\mu -  b^\mu x^2}{1 - 2b_\mu x^\mu + x^2 b^2}$ \,.
\end{enumerate}
It can be shown that these transformations form a group with respect to composition, which is isomorphic to $O(p+1,q+1)/\{\pm 1\}$. Notably, since there are SCT's that are singular (taking points $x\in \mathbb{R}^{p,q}$ to $\infty \notin \mathbb{R}^{p,q}$), these maps are not well defined endomorphisms of $\mathbb{R}^{p,q}$ and thus the introduction of a \textit{conformal compactification} of $\mathbb{R}^{p,q}$ is required. We do not intend to go into detail with respect to the last statement, but we will refer to some ideas to build some intuition for the proceeding sections. Informally, a conformal compactification $\overline{\mathbb{R}^{p,q}}$ of $\mathbb{R}^{p,q}$ is a "minimal" compact set that contains $\mathbb{R}^{p,q}$ where conformal transformations are everywhere defined. The relation between the conformal compactification and the original space is expressed in terms of "good" continuation properties of conformal mappings defined over open subsets $U\subseteq \mathbb{R}^{p,q}$ to mappings defined over $\overline{\mathbb{R}^{p,q}}$ (c.f. \cite{Schottenloher} chapter 2 for further detail). In terms of the conformal compactification, we may define the \textit{conformal Lie group} ${\rm Conf}(\mathbb{R}^{p,q})$ as the connected component containing the identity\footnote{Note that in the same way than with Lorentz transformations, we define the conformal group  in terms of the connected component containing the identity.} of the group of conformal transformations defined over the conformal compactification of $\mathbb{R}^{p,q}$. This group is isomorphic to $SO(p+1,q+1)$ if either $p$ or $q$ are even, and to $SO(p+1,q+1)/\{\pm1\}$ if both $p$ and $q$ are odd. In particular, the conformal Lie algebra we introduced above is the Lie algebra of the conformal Lie group:
\begin{equation}
   \mathrm{Lie } \mathrm{Conf}(\mathbb{R}^{p,q})= {\rm conf}(\mathbb{R}^{p,q})\cong \frak{so}(p+1,q+1) \,. 
\end{equation}

\subsection{Conformal Symmetry in $d=2$}
\label{sec:francescodeq2}

In the previous subsections we discussed conformal symmetry in $d\neq 2$ dimensions. Here we will talk about conformal symmetry in two dimensions which requires special attention. We will see that the whole algebraic structure in the two dimensional case can be understood from a completely different perspective using tools of complex analysis. In particular, we will see the emergence of an infinite dimensional symmetry algebra corresponding to local conformal transformations. Once again, we will present the topic as it is usually done in textbooks leaving the mathematical discussion for the next subsection. 

To study conformal transformations in two dimensions, we first have to choose whether we will be working on Euclidean space $\mathbb{R}^{2,0}$ or Minkowski space $\mathbb{R}^{1,1}$. The standard procedure is to work in Euclidean coordinates under the assumption that the Minkowski case is completely analogous. This is not entirely true, especially at the level of the Lie groups as we will comment in the proceeding section. However, for the Lie algebras, in which we are ultimately interested, the output of either case is very similar.  

Hence, we consider the Euclidean space with coordinates $(x^0,x^1)$ and a coordinate transformation $(x^0,x^1)\mapsto (w^0(x),w^1(x))$. The definition of a conformal transformation given by equations \eqref{ec:metrictransformation} and \eqref{ec:conformalcondition} become a set of differential equations for $w^0(x)$ and $w^1(x)$ given by:
\begin{align*}
      \left(\frac{\partial w^0}{\partial x^0}\right)^2 + \left(\frac{\partial w^0}{\partial x^1}\right)^2 &= \left(\frac{\partial w^1}{\partial x^0}\right)^2 + \left(\frac{\partial w^1}{\partial x^1}\right)^2 \\
      \\
     \frac{\partial w^0}{\partial x^0}\frac{\partial w^1}{\partial x^0} &+ \frac{\partial w^0}{\partial x^1}\frac{\partial w^1}{\partial x^1}  = 0 \,.
\end{align*}
These conditions are equivalent either to:
\begin{equation}
\label{ec:holomorpic}
    \frac{\partial w^1}{\partial x^0}=\frac{\partial w^0}{\partial x^1} \quad \text{and} \quad \frac{\partial w^0}{\partial x^0}=-\frac{\partial w^1}{\partial x^1}
\end{equation}
or to:
\begin{equation}
\label{ec:antiholomorphic}
    \frac{\partial w^1}{\partial x^0}=-\frac{\partial w^0}{\partial x^1} \quad \text{and} \quad \frac{\partial w^0}{\partial x^0}=\frac{\partial w^1}{\partial x^1} \,, 
\end{equation}
from where we can immediately recognize the Cauchy-Riemann conditions for holomorphic functions in equation \eqref{ec:holomorpic} and similarly, for anti holomorphic functions in equation \eqref{ec:antiholomorphic}. This motivates the definition of complex coordinates
\begin{equation}
\label{ec:complexcoordinates1}
\begin{cases}
\begin{array}{l}
     z=x^0+ix^1  \\
     \partial_z
     =\frac{1}{2}(\partial_{0}-i\partial_{1})
\end{array}        
\end{cases}
      \quad \text{and }\quad
\begin{cases}
 \begin{array}{l}
     \bar{z}=x^0-ix^1  \\
     \partial_{\bar{z}}=\frac{1}{2}(\partial_{0}+i\partial_{1}) 
\end{array}        
\end{cases}
\end{equation}
in terms of which the metric tensor becomes:
\begin{equation}
g_{\mu \nu} =
\begin{pmatrix}
0 & 1/2\\
1/2 & 0
\end{pmatrix} \quad \text{and} \quad 
g^{\mu \nu}=\begin{pmatrix}
0 & 2\\
2 & 0
\end{pmatrix} \,.
\end{equation}
In terms of the complex coordinates, the Cauchy-Riemann equations given in \eqref{ec:holomorpic} and \eqref{ec:antiholomorphic} become
\begin{equation}
    \partial_{\bar{z}} w(z,\bar{z})= 0 \quad \text{and} \quad \partial_{z}\bar{w}(z,\bar{z})=0\,.
\end{equation}
This implies that any coordinate transformation given by a holomorphic and antiholomorphic function of $z$ and $\bar{z}$ respectively, will be a conformal transformation: 
\begin{equation}
    z \mapsto w(z) \quad \text{and} \quad \bar{z}\mapsto \bar{w}(\bar{z})\,.
\end{equation}
Hence, the conformal group\footnote{This statement is in fact, ambiguous. See \S \ref{sec:mathematicalaspectsd2} for a detailed discussion.} in two dimensions is the set of all analytic maps over the complex plane. Next, we wish to study infinitesimal transformations, and to that end, we note that any infinitesimal transformation may be written as:
\begin{equation}
\label{ec:francescoinfinitesimal}
    z \mapsto z'=z+\epsilon (z)\,,\quad \bar{z}\mapsto \bar{z}'=\bar{z}+\bar{\epsilon}(\bar{z})\,.
\end{equation}
where the only requirement for these transformations to be conformal is that $\epsilon(z)$ and $\bar{\epsilon}(\bar{z})$ are holomorphic and anti-holomorphic functions of $z$ and $\bar{z}$ respectively. Hence, both $\epsilon(z)$ and $\bar{\epsilon}(\bar{z})$ admit a well defined Laurent expansion:
\begin{equation}
\label{ec:epsilonexpansion}
\begin{array}{c}
z^{\prime}=z+\epsilon(z)=z+\displaystyle\sum_{n \in \mathbb{Z}} \epsilon_{n} z^{n+1}, \\ \\
\bar{z}^{\prime}=\bar{z}+\bar{\epsilon}(\bar{z})=\bar{z}+\displaystyle\sum_{n \in \mathbb{Z}} \bar{\epsilon}_{n}\bar{z}^{n+1},
\end{array}
\end{equation}
where the infinitesimal parameters $\epsilon_n$ and $\bar{\epsilon}_n$ are constant. The generators of these infinitesimal transformations are thus infinitely many, given by:
\begin{equation}
\label{ec:wittgenerators}
    \ell_n = z^{n+1}\partial_z \, \quad \text{and} \quad \bar{\ell}_n=\bar{z}^{n+1}\partial_{\bar{z}}\, \quad \text{with} \quad n\in \mathbb{Z} \,,
\end{equation}
and they satisfy the following commutation relations:
\begin{equation}
\label{ec:wittCR}
    [\ell_m,\ell_n]=(m-n)\ell_{m+n}\,,\quad [\bar{\ell}_m,\bar{\ell}_n]=(m-n)\bar{\ell}_{m+n}\,,\quad [\ell_m,\bar{\ell}_n]=0\,.
\end{equation}
Hence, we conclude that the Lie algebra\footnote{See \S \ref{sec:mathematicalaspectsd2} for a discussion regarding the Lie algebra of  local conformal transformations.} of local conformal transformations in two dimensions is the direct sum of two isomorphic infinite dimensional Lie algebras, known in the literature as the Witt algebra. In particular, we note that each copy of the Witt algebra contains a finite dimensional subalgebra generated by $\{\ell_{\pm 1},\ell_{0}\}$ and $\{\bar{\ell}_{\pm 1},\bar{\ell}_{0}\}$ respectively, which satisfy the commutation relations of an $\frak{sl}(2,\mathbb{R})$ algebra. Moreover, taking into consideration the fact that $\frak{sl}(2,\mathbb{R})\oplus \frak{sl}(2,\mathbb{R})\cong \frak{sl}(2,\mathbb{C})\cong \frak{so}(3,1)$ we find that the finite dimensional subalgebra spanned by $\{\ell_{\pm 1},\ell_{0}\}$ and $\{\bar{\ell}_{\pm 1},\bar{\ell}_{0}\}$ can be identified with the Lie algebra of global conformal transformations\footnote{Above we said that the group of global conformal transformations in two dimensions is the set of all analytic maps over the complex plane. This would be incompatible with the algebra of global conformal transformations being finite dimensional. See \S \ref{sec:mathematicalaspectsd2} section for a discussion on the global conformal group.}. 

Remarkably, we observe that in the two dimensional case, the distinction\footnote{See \S \ref{sec:mathematicalaspectsd2} for a detailed explanation.} between global and local transformations is of fundamental importance given that the latter is responsible for the appearance of an infinite dimensional symmetry algebra. Indeed, it is the infinite dimensionality of the local conformal algebra that makes two dimensional conformal field theories so prosperous as we will discuss in \S \ref{sec:2dCFT}.  

\subsection{Mathematical Aspects in $d=2$}
\label{sec:mathematicalaspectsd2}
Conformal transformations in two dimensions require a more careful treatment than the $d>2$ case. From an intuitive point of view, for $d=1$ every coordinate transformation is conformal\footnote{This can be seen immediately from equation \eqref{ec:conformal2} where for $d=1$ there is no constraint on $f$ so that any coordinate transformation is conformal.}. On the other hand, we have seen that for $d\geq3$ this is not true and thus the two dimensional case appears as a transition scenario. This idea is mainly captured through Liouville's rigidity theorem \cite{Liouville} that states that if $d\geq 3$, then all \textit{local} conformal transformations of $\overline{\mathbb{R}^{p,q}}$ can be extended to \textit{global} conformal transformations. This is the reason why in $d\geq 3$ we didn't distinguish between them. However, in the two dimensional case, not every local conformal transformation may be extended to a global one, and thus we have to study local and global transformations separately. As a matter of fact, local conformal transformations are responsible for the appearance of the infinite dimensional symmetry algebras, which is one of the distinctive features of two dimensional conformal field theories. 

Once again, the use of technical language will be required, in particular because of the subtle differences between the Euclidean and Minkowski spaces, which we will treat separately. In the same way than we did in \S \ref{sec:mathperspective1}, we outline some of the main ideas. In Euclidean spacetime a reasonable notion of a global conformal group can be defined; which in fact, is isomorphic to $SO(3,1)$ as one would expect from the $d\geq 3$ case. When considering local transformations in terms of vector fields, one finds a tensor product of Lie \textit{algebroid} structures, each of which, contains a copy of a Witt algebra, which we then take to be our local symmetry algebra. In Minkowski spacetime, the set of global conformal transformations forms an infinite dimensional group. The associated Lie algebra is again infinite dimensional and contains two copies of the Witt algebra as well. Thus, we see that the discussion in \S \ref{sec:francescodeq2} contains a mixture of some of these ideas, which we intend to make more precise in what follows. 

\subsubsection{Euclidean Space $\mathbb{R}^{2,0}$}
\label{sec:euclideanspace}

The first main result follows in the same way than in the previous section, from equations \eqref{ec:holomorpic} and \eqref{ec:antiholomorphic}. Concretely we have that for $U \subset \mathbb{R}^{2,0}$ open and connected, every conformal (and orientation preserving) transformation $\phi:U\to \mathbb{R}^{2,0}\cong \mathbb{C}$ is locally\footnote{Note that we are considering transformations defined on connected open subsets and not on the whole $\mathbb{R}^{2,0}$.} holomorphic or anti-holomorphic. If we were to define the conformal group ${\rm Conf}(\mathbb{R}^{2,0})$ following the procedure outlined at the end of section \ref{sec:mathperspective1} we should consider the conformal compactification of $\mathbb{R}^{2,0}$. Although this is perfectly possible, the problem appears when considering the "good" continuation properties of the conformal mappings defined over open subsets $U \subset \mathbb{R}^{2,0}$. To be precise, to have a well defined conformal continuation $\hat{\phi}:\overline{\mathbb{R}^{2,0}}\to \overline{\mathbb{R}^{2,0}}$ of our original $\phi$, we require $\phi$ to be injective\footnote{c.f. \cite{Schottenloher} for further detail.}. This condition fails drastically given that there are many non injective holomorphic (and therefore locally conformal) transformations, such as $z\mapsto z^k$ with $k\notin \{\pm 1,0\}$. This is a clear example of Liouville's rigidity theorem, where we note that there are many local conformal transformations on $\mathbb{R}^{2,0}$ that do not admit a well defined conformal continuation to its conformal compactification $\overline{\mathbb{R}^{2,0}}$. Notably, this is not true in $d\geq 3$, since every conformal transformation defined on a connected open subset $U \subseteq \mathbb{R}^{p,q}$ is injective \cite{Schottenloher}. At this point we have two interesting possibilities. The first, is to restrict the set of conformal transformations under consideration, and to admit only injective mappings. This will lead to the definition of the \textit{global conformal group} in two dimensions. On the other hand, we may continue to study the algebraic structure of local conformal transformations This, instead, will lead us to the \textit{local conformal grupoid} and the \textit{local conformal algebroid} respectively. In fact, as we will discuss below, the latter contains a copy of the Witt algebra.

Let us start by defining a \textit{global conformal transformation} as an injective conformal transformation defined over the entire plane $\mathbb{R}^{2,0} \cong \mathbb{C}$ with at most one singular point. Under these conditions we have that any global conformal transformation has a unique conformal continuation defined on the conformal compactification $\overline{\mathbb{R}^{2,0}}$ such that the set of (conformally continued) diffeomorphisms $\psi:\overline{\mathbb{R}^{2,0}}\to \overline{\mathbb{R}^{2,0}}$ form a group with respect to composition. This group is isomorphic to $O(3,1)/\{\pm 1\}$ and the connected component containing the identity is isomorphic to $SO(3,1)$. We refer to the latter as the global conformal group group ${\rm Conf}(\mathbb{R}^{2,0})\cong SO(3,1)$. Moreover, we may define the global conformal algebra ${\rm conf}(\mathbb{R}^{2,0})$ to be the Lie algebra associated to ${\rm Conf}(\mathbb{R}^{2,0})$. Of course, we have that ${\rm conf}(\mathbb{R}^{2,0})\cong \frak{so}(3,1)$. Notably, in this process, we first discarded many conformal transformations that were well defined locally, and only then we made implicit reference to infinitesimal transformations when defining the global conformal algebra. However, infinitesimal transformations only make sense in terms of a local definition and thus the restriction to globally defined tranformations is not strictly necessary when analyzing infinitesimal transformations. 

This takes us to our second alternative, which is to analyze the algebraic structure of locally defined conformal transformations. As we mentioned already, these are the locally defined holomorphic or anti holomorphic mappings over the complex plane. Even though we will not consider conformal continuations for the reasons explained above, it is customary to consider conformal mappings defined on the conformal compactification of $\mathbb{R}^{2,0}$ which is isomorphic to the Riemann sphere $\mathbb{S}^2 = \mathbb{C}\cup \{\infty\}$. The set of locally defined holomorphic mappings over the Riemann sphere, together with composition, form a \textit{grupoid}\footnote{A grupoid is a generalization of a group, in which the binary operation is not neccessarily defined for every element of the set.}. As we said before, we are ultimately interested in studying infinitesimal transformations, and thus, in the same way than in section \ref{sec:mathperspective1}, we consider conformal Killing vector fields. From equations \eqref{ec:killing2d} for $d=2$ and \eqref{ec:confvectorfield}, it is straight forward to verify that every conformal Killing vector field $\xi=(u,v):U\to \mathbb{C}$ defined on a connected open subset $U\subset M$ with conformal Killing factor $\kappa$ satisfies $\partial_\mu \partial^\mu \kappa =0$ as well as $\partial_y u + \partial_x v =0$ and $\partial_x u = \tfrac{1}{2}\kappa = \partial_y v$. Thus, the components of $\xi$ satisfy the Cauchy-Riemann equations and are hence given by either (locally defined) holomorphic or anti-holomorphic functions. The set of locally defined holomorphic (anti-holomorphic) vector fields forms a Lie \textit{algebroid}\footnote{Again, a Lie algebroid is a generalization of the Lie algebra} with the usual Lie bracket between vector fields, that contains a complex Witt algebra as a Lie subalgebra. As in the previous section, the generators of each copy of the Witt algebra may be defined as in equation \eqref{ec:wittgenerators}, satisfying the commutation relations of equation \eqref{ec:wittCR}. We then take our symmetry algebra of local conformal transformations to be the tensor product of the two copies of the Witt algebra.

\subsubsection{Minkowski Space $\mathbb{R}^{1,1}$}

The situation in Minkowski space is slightly different and we will go over it to clarify the differences but also to point out the similarities with $\mathbb{R}^{2,0}$. To begin with, let us start with the following result: A smooth map $\phi=(u,v):U \to \mathbb{R}^{1,1}$ on a connected open subset $U\subset \mathbb{R}^{1,1}$ is conformal if and only if:
\begin{equation}
    \left(\partial_x u\right)^2> \left(\partial_x v\right)^2 \quad \text{and} \quad
    \begin{cases}
        \begin{split}
      &\partial_x u=\partial_y v  \\
      &\partial_y u= \partial_x v
    \end{split}
    \end{cases}
    \quad \text{or} \quad
\begin{cases}
        \begin{split}
      &\partial_x u=-\partial_y v  \\
      &\partial_y u= -\partial_x v
    \end{split}
    \end{cases}
\end{equation}
which follows from equation \eqref{ec:schotenconformal} with $g_{\mu \nu}={\rm diag}(-1,1)$. Remarkably, differently than in $\mathbb{R}^{2,0}$ conformal transformations in $\mathbb{R}^{1,1}$ can be defined globally without any restriction, taking the open and connected subset $U=\mathbb{R}^{1,1}$, which can be shown using light-cone coordinates as done in \textbf{Thm 2.14} in \cite{Schottenloher}. Furthermore, the group of globally defined conformal transformations over Minkowski space is given by \cite{Schottenloher}:
\begin{equation}
 ({\rm Diff}_+(\mathbb{R})\times{\rm Diff}_+(\mathbb{R}))\cup ({\rm Diff}_-(\mathbb{R})\times {\rm Diff}_-(\mathbb{R}))
\end{equation}
where ${\rm Diff}_\pm(\mathbb{R})$ is the infinite dimensional group of orientation preserving (reversing) diffeomorphisms . Notably, since conformal transformations over $\mathbb{R}^{1,1}$ are globally defined, there is no need to consider any kind of conformal continuation, and we could thus define $\mathrm{Conf}(\mathbb{R}^{1,1})$ as the connected component containing the identity of the group of conformal transformations. It is customary however, to replace $\mathbb{R}^{1,1}$ with its conformal compactification $\mathbb{S}^{1,1}$ to work with compact manifolds. With this choice, we define the conformal group in Minkowski spacetime ${\rm Conf}(\mathbb{S}^{1,1})$ as the connected component containing the identity of the group of conformal transformations over $\mathbb{S}^{1,1}$, given by\footnote{Note that we discard the connected component ${\rm Diff}_-(\mathbb{S})\times{\rm Diff}_-(\mathbb{S})$ which does not contain the identity.}:
\begin{equation}
\label{ec:Minkowskiconfgroup}
  {\rm Conf}(\mathbb{R}^{1,1})\cong {\rm Diff}_+(\mathbb{S})\times{\rm Diff}_+(\mathbb{S})\,. 
\end{equation}
We thus find that the group of conformal transformations defined over two dimensional Minkowski space is indeed infinite dimensional. Next, as usual, we are interested in the infinitesimal transformations. The upshot of Minkowski spacetime is that the conformal group as defined in equation \eqref{ec:Minkowskiconfgroup} is in fact a Lie group. Hence, we may consider the associated Lie algebra ${\rm diff}_+(\mathbb{S})$ which is the algebra of smooth vector fields ${\rm Vect}(\mathbb{S})$ (c.f. \cite{Schottenloher} sec 5.1). The complexification of this algebra contains the Witt algebra as a Lie subalgebra, and thus, the complexification of the conformal algebra of two dimensional Minkowski space contains two copies of the Witt algebra, which again, we take as the symmety algebra of local conformal transformations.

To summarize, in this section we have discussed the different algebraic structures appearing when studying conformal symmetry in two dimensions. We have pointed out subtle differences between Euclidean and Minkowski signatures, and in both cases, we outlined the ideas behind the appearance of the two copies of the Witt algebra. In the former, as a Lie subalgebra of the algebroid of locally defined holomorphic conformal Killing vector fields, while in the latter, as a Lie subalgebra of the complexification of the algebra of smooth vector fields defined on the circle $\mathbb{S}$. In either case, we reached a similar conclusion than in the previous section, where we followed \cite{Francesco}.

\section{Conformal Field Theories in $d\geq 3$}
\label{sec:3dCFT}

In the previous section we introduced conformal transformations as a subset of the set of smooth transformations from the spacetime $\mathbb{R}^{p,q}$ to itself. In other words, as a special subset of transformations of the coordinates. Notably, the main object of study in quantum field theories are not the coordinates, but instead, the fields. Hence, in this section we will initiate our analysis on QFT's with conformal invariance, known as conformal field theories. 

Just as when we were exclusively studying conformal transformations, quantum field theories defined in $d\geq 3$ and $d=2$ dimensions are considerably different and we will therefore present them separately. In particular, given that the global conformal group in two dimensions as introduced in \S \ref{sec:euclideanspace}, has the same structure than the conformal group in $d\geq 3$\footnote {Recall that $\mathrm{Conf}(\mathbb{R}^{p,q})\cong SO(p+1,q+1)$ if $p$ or $q$ are even and $\mathrm{Conf}(\mathbb{R}^{2,0})\cong SO(3,1)$.} all of the discussion in this section will hold for $d=2$ as well. However, the converse is not true: The study of CFT's in two dimensions will prove to be significantly more substantial, as we will discuss in the proceeding section.

\subsection{Conformal Invariance in $d\geq 3$}
To discuss conformal invariance of a quantum field theory we should begin by discussing the transformation properties of the fields \cite{salam,Schroer}. As usual, we look for objects with well defined transformation properties under the symmetry group of the system, which in this case is the conformal group. In other words, we look for fields that transform in irreducible representations of $SO(p+1,q+1)$. These are the \textit{quasi-primary fields} which will play a fundamental role in the description of QFT's with conformal invariance. 

\subsubsection{Quasi-Primary Fields}
\label{sec:quasiprimaryfields}

Let us consider a conformal transformation of the coordinates $x\mapsto x'$. As discussed in the previous section, any conformal transformation is a composition of a Lorentz transformation\footnote{Note that we are actually considering transformations belonging to the conformal group as defined in the end of section \ref{sec:mathperspective1}}, a dilatation and a special conformal transformation. Thus, to state the transformation properties of the field, we can treat each of the transformations separately. As usual, we demand our fields to transform in irreducible representations of the Lorentz group. That is, under a Lorentz transformation of the coordinates $x\mapsto x'=\Lambda x$ with $\Lambda \in SO(p,q)$, we require
\begin{equation}
\label{ec:primarylorentz}
    \phi(x)\to \phi'(x')=\mathcal{R}(\Lambda)\phi(x)
\end{equation}
with $\mathcal{R}$ some irreducible representation of $SO(p,q)$. The novelty comes when we consider dilatations and special conformal transformations: If we want our field to transform in an irreducible representation of the full conformal group, compatible with equation \eqref{ec:primarylorentz}, under a dilatation or a special conformal transformation, the field must\footnote{See for instance the classic paper \cite{salam} for a proof of this statement.} transform as:
\begin{equation}
\label{ec:primarydila}
    \phi(x) \to \phi'(x')=\left|\frac{\partial x'}{\partial x}\right|^{-\Delta/d}\phi(x)\,,
\end{equation}
where $\Delta$ is called the \textit{scaling dimension} of the field, which coincides with its \textit{mass dimension}. The Jacobian of the transformation, is given by:
\begin{equation}
    \left|\frac{\partial x'}{\partial x}\right|=\Omega(x)^{-d}\,,
\end{equation}
where $\Omega(x)$ is the conformal factor of the conformal transformation given by:
\begin{equation}
\label{ec:formasomega}
\begin{split}
    &\Omega(x)=\lambda^{d/2} \quad \text{for a dilatation}    \\
    &\Omega(x)=(1-2b\cdot x +b^2 x^2) \quad \text{for a SCT.}
\end{split}
\end{equation}
A field with these transformation properties is called a \textit{quasi-primary field}, and belongs to an irreducible representation of the conformal group. For a general conformal transformation, the transformation of a quasi-primary field can be obtained by combining equations \eqref{ec:primarylorentz} and \eqref{ec:primarydila}. Naturally, not every field appearing in a conformal field theory will transform as a quasi-primary field. For instance, it is straightforward to verify that if $\phi$ is a quasi-primary field then $\partial_\mu \phi$ will necessarily not be a quasi-primary. 

\subsubsection{Covariance of Correlation Functions}
\label{sec:covariancedgeq3}
Having introduced quasi-primary fields as the fields transforming in irreducible representations of the conformal group, we are ready to discuss the transformation properties of correlation functions: After all, these are the physically measurable quantities. So let us consider the correlation function of a product of $n$ quasi-primary fields $\phi_i$ of scaling dimensions $\Delta_i$, which we take to be Lorentz scalars for simplicity. If our theory is conformally invariant at the quantum level, then both the action and the measure should be invariant under conformal transformations, so that:
\begin{equation}
\label{ec:corrfunctiontransf}
\begin{split}
    \langle \phi_1(x_1')\dots \phi_n(x_n')\rangle
    &=\frac{1}{\mathcal{Z}}\int \mathcal{D}\phi\, \phi_1(x_1')\dots \phi_n(x_n')\,e^{-S[\phi]}\\
    &=\frac{1}{\mathcal{Z}}\int \mathcal{D}\phi'\, \phi'_1(x_1')\dots \phi'_n(x_n')\,e^{-S[\phi']}\\
    &=\left|\frac{\partial x'}{\partial x}\right|_{x=x_1}^{-\Delta_1/d} \cdot \dots \cdot \left|\frac{\partial x'}{\partial x}\right|_{x=x_n}^{-\Delta_n/d}\frac{1}{\mathcal{Z}}\int \mathcal{D}\phi\, \phi_1(x_1)\dots \phi_n(x_n)\,e^{-S[\phi]}\\
    &=\left|\frac{\partial x'}{\partial x}\right|_{x=x_1}^{-\Delta_1/d} \cdot \dots \cdot \left|\frac{\partial x'}{\partial x}\right|_{x=x_n}^{-\Delta_n/d}\langle \phi_1(x_1)\dots \phi_n(x_n) \rangle
\end{split}
\end{equation}
where in the second line we simply renamed the integration variable $\phi \to \phi'$, and in the third line we used the transformation properties of quasi-primary fields given in equation \eqref{ec:primarydila}, together with the invariance of both the measure and the action. It is worth noting that equation \eqref{ec:corrfunctiontransf} relates the fields $\phi_i$ evaluated at $x_i$ and $x_i'$ respectively. Under rearrangement, we get the identity: 
\begin{equation}
\label{ec:covariancecondition}
    \langle \phi_1(x_n)\cdot \dots \cdot \phi_n(x_n)\rangle 
   = \left|\frac{\partial x'}{\partial x}\right|_{x=x_1}^{\Delta_1/d} \cdot \dots \cdot \left|\frac{\partial x'}{\partial x}\right|_{x=x_n}^{\Delta_n/d} \langle \phi_1(x'_1)\cdot \dots \cdot \phi_n(x'_n)\rangle \,.
\end{equation}
The covariance condition given in equation \eqref{ec:covariancecondition} imposes very strong constraints on the structure of correlation functions. For instance, if we consider equation \eqref{ec:covariancecondition} for a pair of quasi-primary fields, specialized to a scale transformation $x \mapsto \lambda x$ we have:
\begin{equation}
\label{ec:scaleinvariance}
    \langle \phi_1(x_1)\phi_2(x_2)\rangle = \lambda^{\Delta_1 + \Delta_2}\langle \phi_1(\lambda x_1)\phi_2(\lambda x_2)\rangle \,.
\end{equation}
On the other hand, rotation and translation invariance imply that the correlation can only depend on $|x_1-x_2|$ and thus, 
\begin{equation}
\label{ec:rotationtranslationinvariance}
     \langle \phi_1(x_1)\phi_2(x_2)\rangle=f(|x_1-x_2|)\,, 
\end{equation}
where $f(x)=\lambda^{\Delta_1+\Delta_2}f(\lambda x)$, which can be seen by replacing \eqref{ec:rotationtranslationinvariance} in both sides of equation \eqref{ec:scaleinvariance}. Now this condition can only be satisfied if 
\begin{equation}
\label{ec:menossct}
    f(|x_1-x_2|) = \frac{C_{12}}{|x_1-x_2|^{\Delta_1 + \Delta_2}} = \langle \phi_1(x_1)\phi_2(x_2)\rangle
\end{equation}
for some constant $C_{12}$. It remains to use covariance under special conformal transformations. Combining equations \eqref{ec:menossct} and \eqref{ec:covariancecondition} we have, 
\begin{equation}
    \frac{C_{12}}{|x_1-x_2|^{\Delta_1 + \Delta_2}}
    = \frac{C_{12}}{\Omega(x_1)^{\Delta_1}\Omega(x_2)^{\Delta_2}}\frac{1}{|x'_1-x'_2|^{\Delta_1 + \Delta_2}}
\end{equation}
where $\Omega(x_i)$ is defined for special conformal transformations in equation \eqref{ec:formasomega}. Finally, it is straightforward to verify that under a special conformal transformation:
\begin{equation}
    |x'_1-x'_2| = \frac{|x_1-x_2|}{\Omega(x_1)^{1/2}\Omega(x_2)^{1/2}}
\end{equation}
and thus, we have:
\begin{equation}
     \frac{C_{12}}{|x_1-x_2|^{\Delta_1 + \Delta_2}}=\frac{C_{12}}{\Omega(x_1)^{\Delta_1}\Omega(x_2)^{\Delta_2}}\frac{\left(\Omega(x_1)\Omega(x_2)\right)^{\frac{\Delta_1+\Delta_2}{2}}}{|x_1-x_2|^{\Delta_1+\Delta_2}} \,.
\end{equation}
This final constraint is only satisfied if $\Delta_1 = \Delta_2$ and therefore, we conclude that the two point correlation function between quasi-primary fields is given by:
\begin{equation}
    \langle \phi_1(x_1)\phi_2(x_2)\rangle = \begin{cases}
    \begin{array}{c}
         \displaystyle\frac{C_{12}}{|x_1-x_2|^{2\Delta_1}}  \quad \text{if} \quad \Delta_1=\Delta_2 \\
         0 \quad \quad \quad \quad  \text{  otherwise}
    \end{array}
    \end{cases}
\end{equation}
A completely analogous analysis may be carried out for three point functions to get:
\begin{equation}
     \langle \phi_1(x_1)\phi_2(x_2)\phi_3(x_3)\rangle=\frac{C_{123}}{x_{12}^{\Delta_1+\Delta_2-\Delta_3}x_{13}^{\Delta_1+\Delta_3-\Delta_2}x_{23}^{\Delta_2+\Delta_3-\Delta_1}}
\end{equation}
where $x_{ij}:=|x_i-x_j|$ and $C_{123}$ some constant coefficient. Thus, we observe the remarkable property that two and three point correlation functions of quasi-primary fields are completely fixed up to some coefficient by conformal symmetry \cite{Polyakov2}! Notably, the fixing of correlation functions solely by the use of conformal symmetry ends with the three point function. For general $n$-point correlation functions with $n\geq 4$, the dynamics of the fields must be taken into account to be able to do the computation. At this point we stop our discussion of conformal field theories in $d\geq 3$. We do so, because this is all that we require to dive into the two dimensional case, which is ultimately what we are interested in. The curious reader may look at \cite{Rychkov2} and references therein for material regarding conformally invariant theories in $d\geq 3$.

\section{Conformal Field Theories in $d=2$}
\label{sec:2dCFT}

All of the discussion in the previous section actually holds for the $d=2$ case as well, but the distinctive feature of two dimensions is that in fact, much more can be said, and that's why we treat it in a separate section. Let us start by reminding the reader why this is the case. As we outlined in the beginning of section \ref{sec:mathematicalaspectsd2}, Liouville's rigidity theorem states that if $d\geq 3$, every local conformal transformation defined on $\overline{\mathbb{R}^{p,q}}$\footnote{Recall that $\overline{\mathbb{R}^{p,q}}$ is the conformal compactification of $\mathbb{R}^{p,q}$; see the end of section \ref{sec:mathperspective1} for further detail.} can be extended to a global conformal transformation. This, however, is not true in $d=2$, where there are many local conformal transformations that can't be extended to a well defined global conformal transformation. To be precise, in section \ref{sec:mathematicalaspectsd2} we have seen that in both Euclidean and Minkowski signatures we may take the Lie algebra of local conformal transformations to be (two copies of) the infinite dimensional Witt algebra, in stark contrast with the finite dimensional Lie algebra $\frak{so}(p+1,q+1)$ of global($=$local) conformal transformations in $d\geq 3$. Notably, when studying quantum field theories, we are interested in the local transformation properties of the fields, since the fields \textit{per se} are taken to be local objects. It is precisely the infinite dimensionality of the local symmetry algebra which makes the study of two dimensional conformal field theories so interesting. 


\subsection{Quantization in Two Dimensions}
\label{sec:2dquantization}

In the previous section we discussed conformally invariant quantum field theories in $d\geq 3$, but we did not make reference to any specific quantization procedure. As a matter of fact, the path integral quantization was implicit in the covariance condition given in equation \eqref{ec:covariancecondition}, but there was no need to introduce any kind of operator formalism given that all of the constraints imposed by \textit{global} conformal symmetry could be encoded through the transformation properties of correlation functions of quasi-primary fields. 

In two dimensions the approach is slightly more abstract, because the usual field theory Lagrangian description is not the most accurate language to deal with the infinite dimensionality of the \textit{local} symmetry algebra. Instead, the strategy will be to define a quantization prescription, which will enable us to introduce a Hilbert space structure that will decompose into irreducible representations of the local symmetry algebra. Correlation functions will be given by vacuum expectation values of time-ordered products of operators, which in turn, will be strongly constrained by the structure of the aforementioned irreducible representations \cite{Kac2}. This procedure is completely analogous to the one used in quantum mechanical systems that enjoy rotational symmetry: After quantization, the Hilbert space splits into irreducible representations of $\frak{su}(2)$ so that the energy spectrum and the angular momentum are determined by analyzing the structure of the irreducible representations of $\frak{su}(2)$. 

Having presented the strategy, we outline the procedure in which it will be implemented. There are several nuances involved in the process so it is important to bear in mind where are we heading to. We will start by analyzing the transformation properties of correlation functions, in analogy to what we did in the $d\geq 3$ case (\S \ref{sec:covariancedgeq3}). There, we introduced quasi-primary fields to be the fields transforming in irreducible representations of the global conformal group. In the two dimensional case, we will be interested in fields transforming in irreducible representations of the local symmetry algebra, which as we discussed in \S \ref{sec:francescodeq2}, is infinite dimensional\footnote{Note that differently than with $d\geq 3$, we are using the local symmetry algebra instead of the global conformal group. The reason we make this choice, is precisely because of it's infinite dimensionality. Furthermore, what enables this option is the fact that fields are local objects, and thus we may only consider their local transformation properties!}. We will refer to these fields as \textit{primary fields}, which will turn out to play a central role because they will provide the link between the field theory description and the algebraic approach discussed above. Notably, the same kind of analysis carried out in \S \ref{sec:covariancedgeq3} regarding the transformation properties of the correlation functions will not be enough to fully exploit the local symmetry algebra, so we will require a different mechanism to relate the symmetries of the quantum system with conserved quantities. This will lead to the introduction of the \textit{Ward Identities}, which capture the spirit of Noether's theorem in the quantum theory. Moreover, we will show that in the particular case of conformal symmetry in two dimensions, these Ward Identities can be taken to have a very special form: They relate the transformation of correlation functions with complex contour integrals of the fields appearing within the correlation function. Schematically, $\delta \langle \phi_1\phi_2 \rangle \sim \oint dz \phi_1\phi_2$. The appearance of these integrals is transcendental, because we may use Cauchy's residue theorem to discard everything that is non-singular from the contour integral. Equivalently, the Ward Identities allow us to synthesize all of the information regarding the transformation of correlation functions, in the singular behavior of the integrand! With this in mind, we look for a device to express the product of fields appearing in the integral in terms of a Laurent series, where the singularities appear in their most suitable format to use the residue theorem. This device is known as the \textit{operator product expansion} (OPE). The next step is subtle, and consists of translating this Laurent series expansion called the OPE into commutator language. In particular, since the OPE's capture all of the symmetry information contained in the Ward identities, the transition to commutator language will provide a method to realize the local symmetry algebra at the quantum level, through it's commutation relations. This can be achieved by introducing \textit{radial quantization}, which as any quantization prescription provides the passage from a field theory description to an operator formalism. In this process, the local symmetry algebra, which we recall from \S \ref{sec:francescodeq2} is given by the Witt algebra, will suffer a slight modification and will be replaced by an extension called the \textit{Virasoro algebra}. With this realization of the symmetry algebra in the quantum theory, we will structure our Hilbert space of states by decomposing it into direct sums of representation spaces of irreducible representations of the Virasoro algebra. Correlation functions will be given by vacuum expectation values of \textit{radially ordered}\footnote{As described in \S \ref{sec:radialordering}, in the context of radial quantization, time ordering is replaced by radial ordering.} product of operators, which in turn, will be expressible in terms of inner products of vectors belonging to these representation spaces. Thus, the study of two dimensional conformal field theories turns out to be intimately related to the representation theory of the Virasoro algebra.

\subsection{Transformation of Correlation Functions}

In this section we will discuss the transformation properties of correlation functions. We will start by defining primary fields as the fields transforming in irreducible representations of the local conformal algebra. Next, we derive an expression for the Ward Identities for a general field theory with a symmetry, which we then specify to the two dimensional case with conformal symmetry. In doing so, we introduce the stress energy tensor which will play an essential role throughout the whole discussion. We conclude this section by introducing the operator product expansion, which we will then use to make contact with the proceeding subsection, where we discuss the quantization procedure. 

\subsubsection{Primary Fields}

In section \ref{sec:quasiprimaryfields} we introduced a quasi-primary field as a field transforming in an irreducible representation of the global conformal group in $d\geq 3$. For the two dimensional case, we may consider fields transforming in irreducible representations of the local conformal algebra, which contains the global conformal algebra as a subalgebra (c.f. \S \ref{sec:francescodeq2}). We will refer to these fields as \textit{primary fields}. Throughout this section we will mainly use complex coordinates as introduced in equation \eqref{ec:complexcoordinates1} which we recall, are defined by:
\begin{equation}
\begin{cases}
\begin{array}{l}
     z=x^0+ix^1  \\
     \partial_z
     =\frac{1}{2}(\partial_{0}-i\partial_{1})
\end{array}        
\end{cases}
      \quad \text{and }\quad
\begin{cases}
 \begin{array}{l}
     \bar{z}=x^0-ix^1  \\
     \partial_{\bar{z}}=\frac{1}{2}(\partial_{0}+i\partial_{1}) 
\end{array}        
\end{cases}
\end{equation}
so that $\phi(x^0,x^1)$ is replaced with $\phi(z,\bar{z})$. Having said this, we first consider quasi-primary fields as defined in the previous section, but now for the two dimensional case. As before, we start by analyzing Lorentz transformations. Notably, in two dimensions, the Lorentz group is isomorphic to $SO(2)\cong U(1)$ whose associated Lie algebra $\frak{so}(2)\cong \frak{u}(1)$ is commutative. This implies that there is no analogue of quantization of angular momentum, which arises in $d\geq 3$ from the non-abelian nature of $SO(d)$. As usual, we consider projective\footnote{See appendix \ref{sec:apendicecentral} for a discussion regarding projective and unitary representations.} representations of the symmetry group, which correspond to unitary representations of its universal covering, which in the case of $SO(2)$ is the real line. Now $\mathbb{R}$ admits a continuum of representations characterized by the eigenvalue $s$ of its only generator $S$, given by:
\begin{equation}
    \mathcal{R}^{(s)}(\theta)=e^{i\theta s} \quad \text{with}\quad \theta \in \mathbb{R}\,.
\end{equation}
Notably, for these representations to be induced by projective representations of $SO(2)$, we must demand that $s=\frac{\phi}{2\pi}+k$ with $\phi$ an arbitrary real constant, and $k\in \mathbb{Z}$ \cite{MacKenzie}. Systems with these peculiar values of spin have been studied in the context of quantum mechanical models with fractional spin and its relation to the spin-statistics theorem \cite{MacKenzie, Plyushchay, Arovas, Jackiw2}. For our present purposes, we may restrict to the case in which $s$ takes either integer or half-integer values, so that under a rotation $z\mapsto w=e^{i\theta}z$, $\bar{z}\mapsto \bar{w}=e^{-i\theta}\bar{z}$, a field with spin $s$ will transform as:
\begin{equation}
\label{ec:rotation2d}
    \phi'(w,\bar{w})=(e^{i\theta})^s\phi(z,\bar{z})\,. 
\end{equation}
  
On the other hand, in the same way than in $d\geq 3$, we have that under a scaling or special conformal transformation $z\mapsto w(z)$ and $\bar{z}\mapsto \bar{w}(\bar{z})$, we require our fields to transform as
\begin{equation}
\label{ec:scaling2d}
\phi'(w,\bar{w})=\left(\frac{\partial w}{\partial z}\right)^{-\Delta/2}\left(\frac{\partial \bar{w}}{\partial \bar{z}}\right)^{-\Delta/2}\phi(z,\bar{z})\,.
\end{equation}
This motivates the definition of the holomorphic and anti holomorphic conformal dimensions
\begin{equation}
    h=\frac{1}{2}(\Delta + s) \quad \text{and} \quad \bar{h}=\frac{1}{2}(\Delta-s) \,,
\end{equation}
such that under an arbitrary global conformal transformation $z\mapsto w(z)$ and $\bar{z}\mapsto \bar{w}(\bar{z})$, a field transforming in an irreducible representation of the global conformal group will transform as:
\begin{equation}
   \label{ec:quasiprimaryfield2d}
    \phi'(w,\bar{w})=\left(\frac{\partial w}{\partial z}\right)^{-h}\left(\frac{\partial \bar{w}}{\partial \bar{z}}\right)^{-\bar{h}}\phi(z,\bar{z})\,. 
\end{equation}
Note that equation \eqref{ec:quasiprimaryfield2d} becomes \eqref{ec:scaling2d} for either a scaling or special conformal transformation, while for a pure rotation, we get \eqref{ec:rotation2d}.

In the two dimensional case, we may further distinguish between local and global transformations: For an infinitesimal conformal transformation $w(z)=z+\epsilon(z)$ and $\bar{w}(\bar{z})=\bar{z}+\bar{\epsilon}(\bar{z})$ as introduced in \ref{sec:francescodeq2}, we have to first order:
\begin{equation}
\label{ec:primaryfield}
\begin{split}
    \delta_{\epsilon,\bar{\epsilon}}\phi
    &=\phi'(w,\bar{w})-\phi(z,\bar{z})  \\
    &=\phi'(z,\bar{z})-\phi(z,\bar{z})+\mathcal{O}(\epsilon^2)\\
    &=(1+\partial_z \epsilon)^{-h}(1+\partial_{\bar{z}}\bar{\epsilon})^{-\bar{h}}\phi(z,\bar{z})-\phi(z,\bar{z})\\
    &=-(h\phi \partial_z \epsilon+ \epsilon \partial_z \phi )-(\bar{h}\phi \partial_{\bar{z}}\bar{\epsilon}+\bar{\epsilon}\partial_{\bar{z}}\phi)\,.
\end{split}
\end{equation}
A field that transforms as in \eqref{ec:primaryfield} under a local conformal transformation is called a \textit{primary field}. Notably, all primary fields are quasi primary fields (given that the local conformal algebra has the global conformal algebra as a subalgebra) but the reverse is not true: A field may transform according to \eqref{ec:quasiprimaryfield2d} under a global conformal transformation but for those transformations only. In the same way than in $d\geq 3$, two and three point functions of primary fields will be completely fixed by conformal symmetry, but primary fields will play a much more fundamental role, given that the representation theory of the local conformal algebra is significantly richer than that of its global counterpart. 

\subsubsection{Two and Three Point Functions}
\label{sec:twopointfunction}

Having introduced primary fields, we may proceed to discuss the transformation properties of correlation functions in two dimensions. In the same way than in $d\geq 3$ (c.f \S \ref{sec:covariancedgeq3}) we have that under a global conformal transformation, the correlation function of $n$ primary fields transforms as:
\begin{equation}
    \langle \phi_1(w_1,\bar{w}_1)\dots \phi_n(w_n,\bar{w}_n) \rangle= \prod_{i=1}^n \left(\frac{\partial w}{\partial z}\right)_{w=w_i}^{-h_i}\left(\frac{\partial \bar{w}}{\partial \bar{z}}\right)_{\bar{w}=\bar{w}_i}^{-\bar{h}_i}\langle \phi_1(z_1,\bar{z}_1)\dots \phi_n(z_n,\bar{z}_n) \rangle \,.
\end{equation}
Once again, this relation fixes the two and three point functions, which in complex coordinates we may write, taking into account the relation $x_{ij}=(z_{ij}\bar{z}_{ij})^{1/2}$ as:
\begin{equation}
\label{ec:twopointfunctionalmostorth}
    \langle \phi_{i}(z_i,\bar{z}_i)\phi_{j}(z_j,\bar{z}_j) \rangle =\frac{C_{ij}}{(z_i-z_j)^{2h}(\bar{z}_i-\bar{z}_j)^{2\bar{h}}} \quad \text{if} \quad 
    \begin{cases}
    \begin{array}{c}
         h_1=h_2=h  \\
         \bar{h}_1=\bar{h}_2=\bar{h} 
    \end{array}
    \end{cases} \,,
\end{equation}
where the two point function vanishes if the fields have different conformal dimension. Similarly, we can use global conformal invariance to fix the three point function, up to an unknown coefficient: 
\begin{equation}
\label{ec:threepointfunction}
\left\langle\phi_{i}\left(x_{i}\right) \phi_{j}\left(x_{j}\right) \phi_{k}\left(x_{k}\right)\right\rangle
=C_{ijk} \frac{1}{z_{ij}^{h_{i}+h_{j}-h_{k}}  z_{jk}^{h_{j}+h_{k}-h_{i}} z_{ik}^{h_{k}+h_{i}-h_{j}}}\cdot
\frac{1}{\bar{z}_{ij}^{\bar{h}_{i}+\bar{h}_{j}-\bar{h}_{k}} \bar{z}_{jk}^{\bar{h}_{j}+\bar{h}_{k}-\bar{h}_{i}} \bar{z}_{ik}^{\bar{h}_{k}+\bar{h}_{i}-\bar{h}_{j}}}\,.
\end{equation}
As in $d\geq 3$ global conformal invariance does not fix the four point functions and beyond. However, we have not yet made reference to the local symmetry algebra, which in fact will further constraint the correlation functions significantly. In order to systematically study these constraints, we introduce the \textit{Ward Identities}. 

\subsubsection{Ward Identities}
Classically, in any QFT, we know that a symmetry of the system implies the existence of a conserved Noether current. Of course, these currents need not to be conserved at the quantum level, but their conservation at the classical level imply certain constraints on correlation functions which are captured by the so known \textit{Ward Identities}. In this section we will derive an expression for a general theory in $d$ dimensions with a symmetry, and only then we will restrict to conformal symmetry in two dimensions.

So let $S[\phi]$ be a classical action and let us consider an arbitrary infinitesimal symmetry transformation of the fields
\begin{equation}
\label{ec:symmetrytransf}
    \phi \mapsto \phi'=\phi +\epsilon\delta \phi\,.
\end{equation} 
Given that the transformation is a symmetry, we have that the variation of the Lagrangian is $\delta L =0$\footnote{Actually, we can be more general and take the variation to be a total derivative, but it won't really modify the argument at all.}. Now if we promote $\epsilon$ to be a function of the spacetime coordinates, the Lagrangian is not longer invariant, and its variation is given by:
\begin{equation}
 \begin{split}
     \delta L 
     &=\frac{\partial L}{\partial\left(\partial_{\mu} \phi\right)} \partial_{\mu}(\epsilon \delta \phi)+\frac{\partial L}{\partial \phi} \epsilon \delta \phi \\
     &=\left(\partial_{\mu} \epsilon\right) \frac{\partial L}{\partial\left(\partial_{\mu} \phi\right)} \delta \phi+\left[\frac{\partial L}{\partial\left(\partial_{\mu} \phi\right)} \partial_{\mu} \delta \phi+\frac{\partial L}{\partial \phi} \delta \phi\right] \epsilon\,.
 \end{split}
\end{equation}
But we know that $\delta L=0$ when $\epsilon$ is constant and thus the term in the square brackets should vanish. Hence, we have that 
\begin{equation}
    \delta L = (\partial_\mu \epsilon) J^\mu  \quad \text{with}\quad J^\mu =\frac{\partial L}{\partial\left(\partial_{\mu} \phi\right)} \delta \phi
\end{equation}
so that the action changes as:
\begin{equation}
\label{ec:deltaese1}
    \delta S = \int d^dx\, \delta L = \int d^dx\, (\partial_\mu \epsilon) J^\mu = - \int d^dx \, \epsilon \,\partial_\mu J^\mu =0 \,.
\end{equation}
This is of course, Noether's theorem and $J^\mu$ is the conserved current associated to the symmetry transformation \eqref{ec:symmetrytransf}. We can go to the quantum theory by considering the Euclidean partition function with a background source associated to the classical action $S[\phi]$:
\begin{equation}
    \mathcal{Z}[\mathcal{J}]=\int \mathcal{D}\phi \,\, \mathrm{exp}\left(-S[\phi]+\int d^dx \,\, \mathcal{J}\phi \right)
\end{equation}
and analyzing the effect of the transformation \eqref{ec:symmetrytransf} on the partition function. As a matter of fact, under this transformation, the partition function remains unmodified given that we may treat $\phi'$ as a dummy integration variable: 
\begin{equation}
\label{ec:partitionfunctioniguales}
    \mathcal{Z}'[\mathcal{J}]=\int \mathcal{D}\phi' \,\, \mathrm{exp}\left(-S[\phi']+\int d^dx \,\, \mathcal{J}\phi' \right)=\mathcal{Z}[\mathcal{J}] \,.
\end{equation}
However, we may also replace $\phi'$ in terms of $\phi$ and $S[\phi']$ with $S[\phi]+\delta S$ to get:
\begin{equation}
\label{ec:partitionfunctionprima}
\begin{split}
    \mathcal{Z}'[\mathcal{J}]
    &=\int \mathcal{D}\phi'\,\, \mathrm{exp}\left(-S[\phi]+\int d^dx\,\, \mathcal{J}\phi\right)\mathrm{exp}\left(-\delta S +\int d^dx \,\,\epsilon \,\mathcal{J}\delta \phi\right)\\
    &=\int \mathcal{D}\phi'\,\, \mathrm{exp}\left(-S[\phi]+\int d^dx\,\, \mathcal{J}\phi\right)\mathrm{exp}\left(-\int d^dx \,\, \epsilon\left(\partial_\mu J^\mu - \mathcal{J}\delta \phi\right)\right)\\
    &\approx \int \mathcal{D}\phi'\,\, \mathrm{exp}\left(-S[\phi]+\int d^dx\,\, \mathcal{J}\phi\right)\left(1-\int d^dx \,\, \epsilon\left(\partial_\mu J^\mu - \mathcal{J}\delta \phi\right)\right)
\end{split}
\end{equation}
where in the second line we used the expression for the variation of $S$ given in equation \eqref{ec:deltaese1}. Next, we assume that the measure is also invariant under the symmetry transformation so $\mathcal{D}\phi'=\mathcal{D}\phi$. Putting together equations \eqref{ec:partitionfunctioniguales} and \eqref{ec:partitionfunctionprima} we get:
\begin{equation}
    \int \mathcal{D}\phi\,\, \mathrm{exp}\left(-S[\phi]+\int d^dx\,\, \mathcal{J}\phi\right)\left(\int d^dx \,\, \epsilon(x)\left(\partial_\mu J^\mu - \mathcal{J}\delta \phi\right)\right)=0
\end{equation}
Since this is true for every $\epsilon(x)$, we may get rid of the integral and obtain an expression for each spacetime point:
\begin{equation}
    \int \mathcal{D}\phi\,\, \mathrm{exp}\left(-S[\phi]+\int d^dx\,\, \mathcal{J}\phi\right)\left(\partial_\mu J^\mu - \mathcal{J}(x)\delta \phi\right)=0\,.
\end{equation}
If we differentiate this expression with respect to the external source $\mathcal{J}(x')$ and then set $\mathcal{J}=0$ we get:
\begin{equation}
    \partial_\mu\langle J^\mu(x)\phi(x')\rangle = \delta(x-x')\langle \delta \phi(x')\rangle\,.
\end{equation}
The general form of the Ward identities is obtained by differentiating with respect to the external source evaluated at different spacetime points followed by setting $\mathcal{J}=0$\footnote{Its worth noting that this expression is only valid in the sense of distributions, that is, when integrated against suitable test functions.}:
\begin{equation}
\label{ec:generalwi}
    \partial_\mu\langle J^\mu (x) \phi(x_1)\dots \phi(x_n)\rangle = \sum_{i=1}^n \delta(x-x_i)\langle \phi(x_1)\dots \delta \phi(x_i)\dots \phi(x_n)\rangle
    \,.
\end{equation}
The expression given in equation \eqref{ec:generalwi} is the Ward identity for a general symmetry transformation in a $d$-dimensional theory. We are interested however in specializing to conformal symmetry in two dimensions. We start by integrating equation \eqref{ec:generalwi} against a function which we take to be constant in a neighbourhood of only one of the points $x_i$ and vanishes elsewhere:
\begin{equation}
\label{ec:wardidentityintegrated}
    \int_C d^dx\,\, \partial_\mu \langle J^\mu (x) \phi(x_1)\dots \rangle = \langle \delta \phi(x_1)\dots \rangle
\end{equation}
where $C$ is the compact region where the constant function is non vanishing which we have taken to be in a neighbourhood of $x_1$ for simplicity. Note that we have replaced $\dots \phi(x_n)$ with $\dots$ given that the test function vanishes in every $x_i\neq x_1$ and so the other fields appearing in the string play no role. Specializing to two dimensions, we may use the divergence theorem to convert the integral in the left hand side of equation \eqref{ec:wardidentityintegrated} into a line integral around the boundary of $C$:
\begin{equation}
\label{ec:lineintegral}
\begin{split}
     \int_C d^2x\,\, \partial_\mu \langle J^\mu (x) \phi(x_1)\dots \rangle 
     &= \oint_{\partial C} \langle J_\mu (x) \phi(x_1)\dots \rangle d \ell^\mu \\
     &
\begin{split}
     =-i\left(\oint_{\partial C}\langle \right.
     & \left. J_z(z,\bar{z}) \phi(z_1,\bar{z}_1)\dots\rangle dz- \right. \\
     &\left. \oint_{\partial C}\langle J_{\bar{z}}(z,\bar{z}) \phi(z_1,\bar{z}_1)\dots\rangle d\bar{z}\right)    
\end{split}     
\end{split}
\end{equation}
where in the first line we introduced the outward-directed differential of circunference  orthogonal to the boundary $\partial C$ of the domain of integration $d\ell^\mu$, and in the second line we have switched to complex coordinates doing the suitable modifications. Putting together equations \eqref{ec:wardidentityintegrated} and \eqref{ec:lineintegral} we get:
\begin{equation}
\label{ec:2dWardIdentity}
    \begin{split}
    \langle \delta \phi(z_1,\bar{z}_1)\dots \rangle=
     -i\left(\oint_{\partial C}\langle \right.
     &\left. J_z(z,\bar{z}) \phi(z_1,\bar{z}_1)\dots\rangle dz- \right. \\
     &\left. \oint_{\partial C}\langle J_{\bar{z}}(z,\bar{z}) \phi(z_1,\bar{z}_1)\dots\rangle d\bar{z}\right)    \,.
\end{split}
\end{equation}
Equation \eqref{ec:2dWardIdentity} is the integrated version of the Ward identities specialized to two dimensions, but we still haven't specified the symmetry: The currents $J_z$ and $J_{\bar{z}}$ are in principle the conserved currents associated to a general symmetry expressed in complex coordinates. To discuss conformal symmetry, it is convenient to introduce the stress energy tensor, in terms of which we will be able to express the conformal currents.

\subsubsection{The Stress Energy Tensor}

The stress energy tensor $T^{\mu \nu}$ is defined to be the Noether current associated to infinitesimal translations:
\begin{equation}
\label{ec:infinitesimaltranslation}
    x^\mu \mapsto x^\mu +\epsilon^\mu \quad \text{with} \quad \mu=1,2
\end{equation}
To find an expression for the stress energy tensor, in the same way than before, we promote $\epsilon \to \epsilon(x)$ in equation \eqref{ec:infinitesimaltranslation} to be a function of the spacetime coordinates. The variation of the action will be thus given by 
\begin{equation}
\label{ec:deltaese}
    \delta S =  \int d^2 x\, ( \partial_\mu \epsilon_\nu) J^{\mu\nu} 
\end{equation}
as in \eqref{ec:deltaese1}. Next, we do a very useful trick \cite{tong}, which is to consider the same theory coupled to a dynamical background $g_{\mu \nu}(x)$. Then, we may regard the infinitesimal transformation
\begin{equation}
    x^\mu \mapsto x^\mu + \epsilon^\mu(x) 
\end{equation}
to be a diffeomorphism, and we know that the theory is invariant as long as we make the corresponding change to the metric
\begin{equation}
    \delta g_{\mu\nu}=\partial_\mu \epsilon_\nu +\partial_\nu \epsilon_\mu \,.
\end{equation}
Thus, if we \textit{only} transform our coordinates in our original theory, the change in the action will be exactly the opposite as if we had \textit{only} transformed the metric, given that doing both should leave the action invariant. Hence, 
\begin{equation}
\label{ec:deltaese2}
    \delta S = -2\int d^2x\, \sqrt{g} \,\frac{\delta S}{\delta g_{\mu\nu}}\partial_\mu\epsilon_\nu\,,
\end{equation}
and therefore, comparing with equation \eqref{ec:deltaese} we obtain an expression for the stress tensor given by:
\begin{equation}
\label{ec:stresstensor}
    T_{\mu \nu}=-\frac{1}{2\sqrt{g}}\frac{\delta S}{\delta g^{\mu \nu}}\,. 
\end{equation}
Under the assumption that the theory is invariant under translations we have, integrating by parts \eqref{ec:deltaese2} that $\partial^\mu T_{\mu \nu}=0$. In theories with conformal symmetry, the stress energy tensor has an additional fundamental property; it is traceless \cite{Luscher}. To see this, we vary our action with respect to a scale transformation, which is a particular example of a conformal transformation,
\begin{equation}
    \delta g_{\mu \nu } = \epsilon g_{\mu \nu}
\end{equation}
and therefore, 
\begin{equation}
    \delta S = \int d^2x\,\sqrt{g}\, \frac{\delta S}{\delta g_{\mu \nu}}\delta g_{\mu \nu} = -2\int d^2x\, \epsilon \, T^{\mu}{}_{\mu} \,.
\end{equation}
But if our theory is conformally invariant, the transformation should be a symmetry of the action, and thus $\delta S=0$, which implies:
\begin{equation}
T^{\mu}{}_{\mu}=0 \,.   
\end{equation}
It is worth noting that the vanishing of the trace holds classically, and although we did the computation in two dimensions, it holds in general. However, this property is not usually preserved when passing to the quantum theory.

Next, we switch back to complex coordinates. The components of the stress tensor are then given by:
\begin{align*}
     T_{zz}&=\frac{1}{4}(T_{00}-2iT_{10}-T_{11})\\
     T_{\bar{z}\bar{z}}&=\frac{1}{4}(T_{00}+2iT_{10}-T_{11})\\
     T_{z\bar{z}}&=T_{\bar{z}z}=\frac{1}{4}(T_{00}+T_{11})=\frac{1}{4}T^{\mu}{}_\mu
\end{align*}
and therefore, traceleness of the stress tensor translates into $T_{z\bar{z}}=T_{\bar{z}z}=0$ while conservation implies $\partial_{\bar{z}} T_{zz}=0$ and $\partial_z T_{\bar{z}\bar{z}}=0$. This motivates the definition of the holomorphic and anti-holomorphic functions given by the non vanishing components of the stress tensor
\begin{equation}
T(z):=T_{zz} \quad \text{and} \quad \bar{T}(\bar{z}):= T_{\bar{z}\bar{z}}\,.
\end{equation} 
Having found the stress tensor, we look for the Noether conserved currents associated to the remaining conformal transformations. In order to do so, we consider a general infinitesimal conformal transformation:
\begin{equation}
\label{ec:inficonftranfo}
    z\mapsto z+\epsilon(z)\,,\quad \bar{z}\mapsto \bar{z}+\bar{\epsilon}(\bar{z})
\end{equation}
with $\epsilon$ and $\bar{\epsilon}$ analytic functions of $z$ and $\bar{z}$ respectively as in equation \eqref{ec:francescoinfinitesimal}. To compute the current, we do the same trick as before so we promote $\epsilon$ to be a function of coordinates. Notably, $\epsilon$ is already $z$ dependent so we take $\epsilon(z)\to \epsilon(z,\bar{z})$ and similarly with $\bar{\epsilon}$. Once again, regarding the transformation as a diffeomorphism, the change in the action will be given by a compensating change in the metric:
\begin{equation}
\label{ec:noethercurrents1}
\begin{split}
    \delta S 
    &= -\int d^2x \, \frac{\delta S}{\delta g^{\mu \nu}}\delta g^{\mu \nu}    \\
    & =\int d^2x \, T_{\mu \nu} \partial^\mu \epsilon^\nu \\
    &= \int d^2z \, (T_{zz}\partial^z \epsilon + T_{z\bar{z}}\partial^{\bar{z}}\bar{\epsilon}+T_{\bar{z}z}\partial^{\bar{z}}\epsilon +T_{\bar{z}\bar{z}}\partial^{\bar{z}}\bar{\epsilon})\\
    &=\int d^2z 
    \,(T_{zz}\partial_{\bar{z}} \epsilon + T_{\bar{z}\bar{z}}\partial_z \bar{\epsilon})\,.
\end{split}
\end{equation}
where in the last line we have used the fact that $T_{z\bar{z}}=T_{\bar{z}z}=0$ and that $\partial^z=g^{z \mu}\partial_\mu=\partial_{\bar{z}}$ and similarly with $\partial^{\bar{z}}=\partial_z$ where $g^{\mu \nu}=\big(\begin{smallmatrix}
  0 & 2\\
  2 & 0
\end{smallmatrix}\big)$. At this point we do another useful trick which is to take $z$ and $\bar{z}$ to be independent variables. In doing so, we are actually extending the range of our coordinates from $\mathbb{R}^2$ to $\mathbb{C}^2$ so it is important to remember that in fact the physical space is the two dimensional submanifold defined by $\bar{z}=z^*$. Under these conditions, we may consider separate transformations for $z$ and $\bar{z}$. Taking $\bar{\epsilon}=0$ and integrating by parts equation \eqref{ec:noethercurrents1} we find that the vanishing of the variation of the action implies the existence of a conserved holomorphic current:
\begin{equation}
\label{ec:holocurrent}
   J^\mu = (J^z,J^{\bar{z}}) \quad \text{with} \quad J^z=0 \quad \text{and} \quad J^{\bar{z}}=T(z)\epsilon (z)\,,
\end{equation}
which satisfies $\partial_\mu J^\mu = 0$. Similarly, taking $\epsilon = 0$ we find the anti-holomorphic current
\begin{equation}
\label{ec:antiholocurrent}
   \bar{J}^\mu = (\bar{J}^z,\bar{J}^{\bar{z}}) \quad \text{with} \quad \bar{J}^z=\bar{T}(\bar{z})\bar{\epsilon}(\bar{z}) \quad \text{and} \quad \bar{J}^{\bar{z}}=0\,.
\end{equation}
which again satisfies the conservation equation $\partial_\mu \bar{J}^\mu=0$. The holomorphic and anti-holomorphic currents defined in equations \eqref{ec:holocurrent} and \eqref{ec:antiholocurrent} are thus the conserved currents associated to a general infinitesimal conformal transformation. 

\subsubsection{Conformal Ward Identity}
\label{sec:conformalwardidentity}

Having found the conserved currents associated to infinitesimal conformal transformations in terms of the components of the stress tensor, we can now specialize equation \eqref{ec:2dWardIdentity} to a conformal symmetry transformation. To make this explicit, we denote by $\delta \phi:=\delta_{\epsilon \bar{\epsilon}} \phi$ the transformation of the fields under the infinitesimal conformal transformation of the coordinates. Once again, we do the trick of treating $z$ and $\bar{z}$ as independent variables so that we may consider the transformations on $z$ and $\bar{z}$ separately. The Ward identity then splits into two pieces; from the change $\delta z = \epsilon(z)$, we get:
\begin{equation}
\label{ec:conformalward1}
    \langle \delta_\epsilon \phi(z_1,\bar{z}_1)\dots \rangle=-\frac{1}{2\pi i} \oint_{C} dz \,\, \epsilon (z)\langle T(z) \phi(z_1,\bar{z}_1)\dots \rangle\,,
\end{equation}
where we have replaced $J_z=J^{\bar{z}}=\epsilon (z)T(z)$ and $J_{\bar{z}}=J^{z}=0$ as in \eqref{ec:holocurrent} and included the $1/2\pi$ factor for convenience. Similarly, for $\delta \bar{z}=\bar{\epsilon}(\bar{z})$ we get:
\begin{equation}
\label{ec:conformalward2}
    \langle \delta_{\bar{\epsilon}} \phi(z_1,\bar{z}_1)\dots \rangle=\frac{1}{2\pi i} \oint_{C} d\bar{z} \,\, \bar{\epsilon} (\bar{z})\langle \bar{T}(\bar{z}) \phi(z_1,\bar{z}_1)\dots \rangle
\end{equation}
where we have replaced the expressions for $\bar{J}_z=\bar{J}^{\bar{z}}=0$ and $\bar{J}_{\bar{z}}=\bar{J}^z=\bar{\epsilon} (\bar{z}) \bar{T}(\bar{z})$ as in \eqref{ec:antiholocurrent}. For a general conformal transformation, we may combine equations \eqref{ec:conformalward1} and \eqref{ec:conformalward2} to write the \textit{conformal Ward identity} \cite{Friedan} as:
\begin{equation}
\label{ec:conformalWI}
\begin{split}
    \langle \delta_{\epsilon \bar{\epsilon}} \phi(z_1,\bar{z}_1)\dots \rangle
    &=-\frac{1}{2\pi i} \oint_{C} dz \,\, \epsilon (z)\langle T(z) \phi(z_1,\bar{z}_1)\dots \rangle \\   
    &+\frac{1}{2\pi i} \oint_{C} d\bar{z} \,\, \bar{\epsilon} (\bar{z})\langle \bar{T}(\bar{z}) \phi(z_1,\bar{z}_1)\dots \rangle\,,
\end{split}
\end{equation}
where the contour of integration only includes the point $(z_1,\bar{z}_1)$, which of course, is arbitrary. Equation \eqref{ec:conformalWI} is the main result of this section.

\subsubsection{Operator Product Expansion}
\label{sec:OPE}

Equation \eqref{ec:conformalWI} may be simplified even further using Cauchy's residue theorem. In order to do so, we must look for a method to Laurent expand the integrand. This can be achieved with an \textit{operator product expansion} (OPE)  which allows us to represent the correlation function of a time ordered product of two local operators inserted at nearby points $z$ and $w$ respectively, by a sum of terms, each being a single operator, well defined as $z\to w$ multiplied by a function, which on grounds of translational invariance depends on $z-w$ \cite{Wilson2, Kadanoff}. These functions, possibly divergent, embody the singular behavior of the operators as their insertion points tend towards each other. Explicitly, given a pair of operators $\mathcal{O}_i$ and $\mathcal{O}_j$ we may write:
\begin{equation}
\label{ec:OPE123}
    \langle \mathcal{O}_i(z,\bar{z})\mathcal{O}_j(w,\bar{w}) \rangle = \sum_{k}C_{ij}^k(z-w,\bar{z}-\bar{w})\langle \mathcal{O}_k(w,\bar{w}) \rangle
\end{equation}
where the sum runs over all of the operators in the theory and the $C_{ij}^k(z-w,\bar{z}-\bar{w})$ are the possibly divergent $c-$valued functions. As a matter of fact, an OPE can be done in any (not necessarily conformal) QFT, but only in the asymptotic short distance limit. The key feature of CFT's is that the OPE is not just asymptotic, it is an exact statement: The series is convergent at finite point separation as we will discuss in \S \ref{sec:asymptoticstates}. 

We may compute our first OPE by looking at the conformal Ward identity \eqref{ec:conformalWI} for a single primary field of conformal dimensions $(h,\bar{h})$, and requiring the transformation to be equal to the one given in equation \eqref{ec:primaryfield}. Using the residue theorem, it is straightforward to verify that the OPE has to take the form:
\begin{equation}
\label{ec:primaryfieldope1}
\begin{split}
 \langle T(z)\phi(w,\bar{w}) \rangle &=\frac{h\langle\phi(w,\bar{w})\rangle}{(z-w)^2}+\frac{\langle\partial\phi(w,\bar{w})\rangle}{z-w} +\mathrm{reg}\\
     \langle\bar{T}(\bar{z})\phi(w,\bar{w}) \rangle &=\frac{\bar{h}\langle\phi(w,\bar{w})\rangle}{(\bar{z}-\bar{w})^2}+\frac{\langle \partial \phi (w,\bar{w})\rangle}{\bar{z}-\bar{w}}+\mathrm{reg}\,.
\end{split}
\end{equation}
where "$\mathrm{reg}$" stands for terms that are regular as $z\to w$. Actually, we are not computing the whole OPE given that we are leaving a (possibly infinite) number of unspecified terms. The point is that we only care about the singular terms of the OPE because these are the ones who actually contain the important information. As a matter of fact, equation \eqref{ec:primaryfieldope1} is exactly equivalent to equation \eqref{ec:primaryfield} and it can thus be alternatively taken to be the definition of a primary field of conformal dimensions $(h,\bar{h})$.

Of course, in order to be able to extract some information from the OPE, we should be able to compute the coefficients $C_{ij}^k$, which for instance, can be calculated using Wick's theorem. The reader may refer to \S 5.3 in \cite{Francesco} where several OPE's are computed for different free theories. Of fundamental importance, is the self OPE of the energy stress tensor of a theory, whose general structure is given by
\begin{equation}
\label{ec:TTselfOPE1}
\begin{split}
\langle T(z)T(w) \rangle
&= \frac{c/2}{(z-w)^4}+\frac{2 \langle T(w) \rangle}{(z-w)^2}+\frac{\langle \partial T(w)\rangle}{z-w} + \mathrm{reg}    \\
\langle \bar{T}(\bar{z})\bar{T}(\bar{w}) \rangle
&= \frac{c/2}{(\bar{z}-\bar{w})^4}+\frac{2 \langle \bar{T}(\bar{w})\rangle}{(\bar{z}-\bar{w})^2}+\frac{\langle\bar{\partial} \bar{T}(\bar{w})\rangle}{\bar{z}-\bar{w}} + \mathrm{reg}\,.
\end{split}    
\end{equation}
The number $c$ appearing in the $(z-w)^{-4}$ term is called the \textit{central charge}, and it's value depends on the theory under consideration; for instance, $c=1$ for the free boson and $c=\tfrac{1}{2}$ for the free fermion (c.f. \S 5.3.1 and 5.3.2 in \cite{Francesco} for the explicit computation). Very importantly, the value of the central charge may not be determined from symmetry considerations; it's value is determined by the short distance behaviour of the theory. There are several reasons and consistency checks to argue that equation \eqref{ec:TTselfOPE1} is the most general form the self OPE of the stress tensor can take. However, there is no actual proof for this statement in general. We will come back to this discussion in \S \ref{sec:virasoroalgebra}. In particular, by replacing the OPEs  given in equation \eqref{ec:TTselfOPE1} in the Ward Identity \eqref{ec:conformalward2}, we recover the transformation relation given in equation \eqref{ec:quasiprimaryfield2d}, and thus we conclude that $T(z)$ and $\bar{T}(\bar{z})$ are quasi-primary fields of conformal dimensions $(2,0)$ and $(0,2)$ respectively. Notably, because of the central charge term in the OPEs, equation \eqref{ec:primaryfield} is not satisfied and therefore, the components of the stress tensor are not primary fields.

Nonetheless, the main takeaway is that the transformation properties of correlation functions can be directly translated into OPE's and viceversa by means of the Ward Identities. Most importantly, in the next section we will introduce a quantization prescription, and we will show how the OPE enables us to realize the local symmetry algebra in the quantum theory.   

\subsection{Radial Quantization}
\label{sec:radialquantization}

In the previous section we derived an expression for the conformal Ward Identity, which describes the transformation properties of a correlation function under a conformal transformation. Notably, up to this point we have discussed the quantum aspects of conformal field theories only in the path integral formulation. In this section we introduce a quantization prescription, \textit{radial quantization}, which will allow us to translate all of these ideas into operator language \cite{Fubini}. We'll see how the introduction of a Hilbert space structure will enable us to exploit the algebraic and group-theoretic methods which in the case of two dimensional conformal field theories become extremely powerful because of the infinite dimensionality of the symmetry algebra. 

\begin{figure}[h]
    \centering
    \includegraphics[scale=0.6]{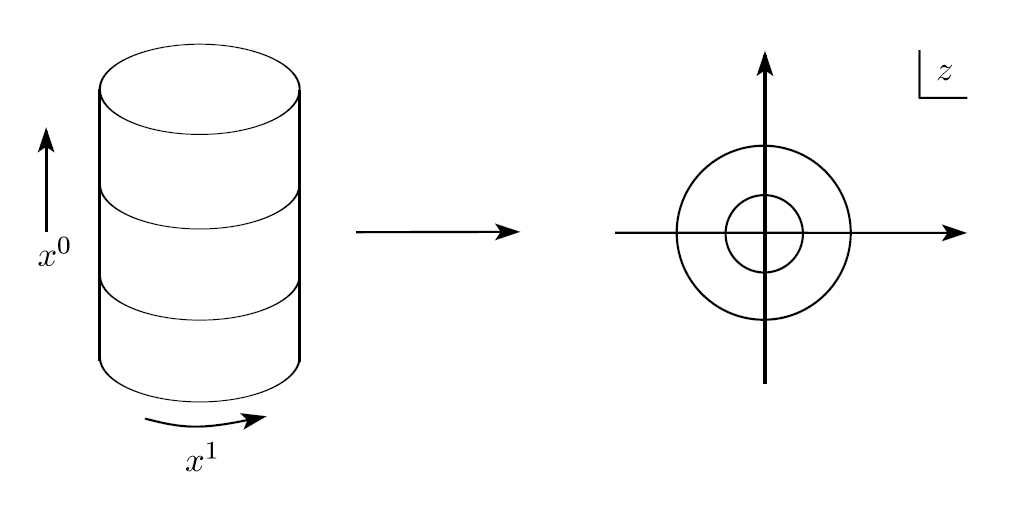}
    \caption{Conformal mapping of the cylinder into the complex plane}
    \label{fig:radialquant}
\end{figure}

We start by considering flat Euclidean space and time coordinates given by $x^0$ and $x^1$ respectively. To eliminate any infrared divergences we compactify the space coordinate such that $x^1 \sim x^1 + 2\pi$, which defines a cylinder in the $(x^0,x^1)$ plane. Next we define the complex coordinate $\zeta=x^0+ix^1$ and we consider the conformal map
\begin{equation}
\label{ec:conformalmap}
    \zeta \mapsto z=e^{\zeta}=e^{x^0+ix^1}
\end{equation}
which maps the cylinder into the complex plane, as in figure \ref{fig:radialquant}. In this way, the infinite past $x^0=-\infty$ is mapped into $z=0$ while the infinite future $x^0=+\infty$ is mapped into $z=\infty$ (in the Riemann sphere). Moreover, equal time surfaces $x^0={\rm constant}$, are mapped into circles of constant radius in the $z$-plane. Hence, we will define a Hilbert space structure in each of the surfaces of constant radius. Notably, time translations in the cylinder become complex dilatations in the plane, and thus the evolution of states will be dictated by the dilatation operator. 

To construct the Hilbert space, we first look for a realization of the symmetry algebra in terms of operators. In the previous section, we have seen that all of the symmetry information is contained in the Ward identity \eqref{ec:conformalWI}, which on turn was expressible in terms of the OPE's. In what follows, we will introduce \textit{radial ordering}, which will enable us to translate OPE's into commutation relations.

\subsubsection{Radial Ordering}
\label{sec:radialordering}

\begin{figure}[h]
    \centering
    \includegraphics[scale=0.5]{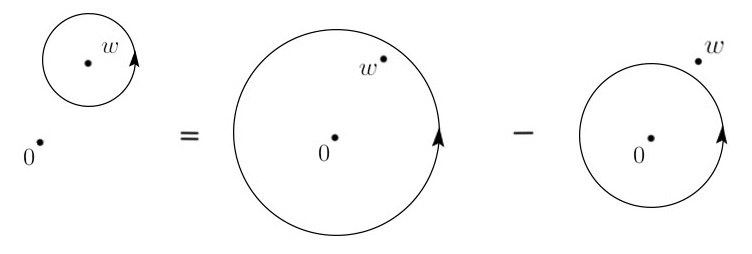}
    \caption{Substraction of contours}
    \label{fig:contoursplit}
\end{figure}

In a quantum field theory, the fields appearing within correlation functions are taken to be time ordered. Notably, when passing from the cylinder to the complex plane, time ordering becomes radial ordering. This means that operators in correlation functions are ordered so that those inserted at larger radial distance are moved to the left. Explicitly, given a pair of operators $a$ and $b$ in the complex plane, we may define the radial ordering operation by:
\begin{equation}
    \mathcal{R}(a(z)b(w))=
    \begin{cases}
    \begin{array}{c}
         a(z)b(w) \quad \text{if} \quad |z|>|w|  \\
         b(w)a(z) \quad \text{if} \quad |w|>|z| \,.
    \end{array}
    \end{cases}
\end{equation}
In the operator formalism, every operator expression appearing within correlation functions will thus be interpreted as to be radially ordered. For simplicity, we will sometimes omit the radial ordering operator $\mathcal{R}$, but radial ordering will always be implicit. We may use radial ordering to relate OPE's with commutation relations as follows. Let $a(z)$ and $b(z)$ be two holomorphic fields and let's consider the integral
\begin{equation}
    \oint_w dz \, \mathcal{R}(a(z)b(w))
\end{equation}
where the integration contour circles counterclockwise around $w$. Next,  employing the relation among contour integrals shown in Fig. \ref{fig:contoursplit} we write our integral as:
\begin{equation}
\label{ec:commutator10}
\begin{split}
      \oint_w dz \, \mathcal{R}(a(z)b(w)) 
      &= \oint_{|z|>|w|}dz \, a(z)b(w) - \oint_{|w|>|z|} dz\, b(w)a(z)  \\
      &= [A,b(w)]
\end{split}
\end{equation}
where we have defined the operator $A$ to be the integral of $a(z)$ at fixed radius (i.e. fixed time)
\begin{equation}
    A=\oint dz\, a(z)\,.
\end{equation}
In practice, we would like to replace the radially ordered product $\mathcal{R}(a(z)b(w))$ appearing in the left hand side of equation \eqref{ec:commutator10} with the corresponding OPE, but there is an important caveat: OPE's are defined within correlation functions! Thus, if we want to use the OPE as an operator identity (i.e. removing the brackets $\langle \dots \rangle$) we should verify that equation \eqref{ec:commutator10} holds when the fields $a(z)$ and $b(w)$ are inserted within a generic correlator. Indeed, for a correlator containing an arbitrary number of additional fields, the contour decomposition given in figure \ref{fig:contoursplit} is valid as long as $b(w)$ is the only other field having a singular OPE with $a(z)$, which lies between the two circles. In fact, we can guarantee this is the case by choosing the contours of integration wisely; if we take the circles to have radii $w+\epsilon$ and $w-\epsilon$ respectively, in the limit $\epsilon\to 0$, the insertion point of any other operator will be excluded from between the circles, thus satisfying the aforementioned condition.  

Hence, we may define the commutator $[A,B]$ of two operators, each the integral of a holomorphic field, by integrating equation \eqref{ec:commutator10} over $w$:
\begin{equation}
\label{ec:OPEcommutator}
    [A,B]=\oint_0 dw \oint_w dz\, \mathcal{R}(a(z)b(w)) \,,
\end{equation}
where
\begin{equation}
    A=\oint dz \, a(z) \quad \text{and} \quad B =\oint dz \, b(z)\,.
\end{equation}
From the above discussion, we note that equation \eqref{ec:OPEcommutator} is valid as an operator expression, so that from now on, we may safely remove the brackets $\langle \dots \rangle$ whenever we replace a radially ordered product with its corresponding OPE. In fact, equation \eqref{ec:OPEcommutator} is of fundamental importance because we may use it to relate OPE's with commutation relations: This is a crucial step in the realization of the symmetry algebra at the quantum level. 

\subsubsection{The Virasoro Algebra}
\label{sec:virasoroalgebra}

We are ready to establish the algebraic structure discussed at the beginning of this section. With all of the tools we have developed, we may now represent the generators of the local symmetry algebra in terms of operators of the quantum theory. This will enable us to decompose our Hilbert space as a direct sum of its representation spaces. 

To begin with, we note that from the preceding discussion regarding radial ordering, that we may rewrite the conformal Ward identity given in \eqref{ec:2dWardIdentity} as a commutator relation:
\begin{equation}
\label{ec:comutatorWI}
        \delta_{\epsilon \bar{\epsilon}}\langle \phi(w,\bar{w})\rangle = -[Q_\epsilon,\phi(w,\bar{w})]+[Q_{\bar{\epsilon}},\phi(w,\bar{w})]
\end{equation}
where we have defined the \textit{conformal charges}:
\begin{equation}
    Q_\epsilon = \frac{1}{2\pi i}\oint_C dz\, \epsilon(z)T(z)\,,\quad Q_{\bar{\epsilon}} = \frac{1}{2\pi i}\oint_C d\bar{z}\, \bar{\epsilon}(\bar{z})\bar{T}(\bar{z})\,.
\end{equation}
Equation \eqref{ec:comutatorWI} implies that the operators $Q_\epsilon$ and $Q_{\bar{\epsilon}}$ are the generators of infinitesimal conformal transformations of the quantum theory. As a matter of fact, we may identify the proper generators of the local symmetry algebra by Laurent expanding the components of the stress energy tensor:
\begin{align}
    T(z)&=\sum_{n\in \mathbb{Z}}z^{-n-2}L_n \quad \text{with} \quad L_n=\frac{1}{2\pi i}\oint_C dz\, z^{n+1}T(z)\,,\\
    \bar{T}(\bar{z})&=\sum_{n\in \mathbb{Z}}\bar{z}^{-n-2}\bar{L}_n \quad \text{with} \quad \bar{L}_n=\frac{1}{2\pi i}\oint_C d\bar{z}\, \bar{z}^{n+1}\bar{T}(\bar{z})\,.
\end{align}
The mode operators $L_n$ and $\bar{L}_n$ will thus be the generators of the local symmetry algebra at the level of the Hilbert space, in the same way than the Witt generators $\ell_n$ and $\bar{\ell}_n$ are the generators of conformal transformations of coordinates (c.f. \S \ref{sec:francescodeq2}). We may compute the commutation relations between the $L_n$ using the self OPE of the stress tensor given in equation \eqref{ec:TTselfOPE1} as follows:
\begin{equation}
\begin{split}
    [L_n,L_m]
    &=\frac{1}{(2\pi i)^2}\oint_0 dw \, w^{m+1}\oint_w dz \, z^{n+1} \left[\frac{c/2}{(z-w)^4}+\frac{2 T(w)}{(z-w)^2}+\frac{\partial T(w)}{z-w} + \mathrm{reg}\right] \\
    &=\frac{1}{2\pi i}\oint_0 dw \, w^{m+1}\left[\frac{1}{12}c(n+1)n(n-1)w^{n-2}+2(n+1)w^n T(w)+ w^{n+1}\partial T(w)\right]\\
    &=\frac{1}{12}cn(n^2-1)\delta_{n+m,0}+2(n+1)L_{n+m} -\frac{1}{2\pi i}\oint_0 dw \, w^{n+m+1}T(w)\\
    &=\frac{1}{12}cn(n^2-1)\delta_{n+m,0}+(n-m)L_{n+m}\,.
    \end{split}
\end{equation}
Similarly, we have that:
\begin{equation}
    [\bar{L}_n,\bar{L}_m]=\frac{1}{12}cn(n^2-1)\delta_{n+m,0}+(n-m)\bar{L}_{n+m} \,.
\end{equation}
and finally,
\begin{equation}
\label{ec:decoupledmodes}
   [L_n,\bar{L}_m]=0 
\end{equation}
which follows from the trivial OPE $\langle T(z)\bar{T}(\bar{w})\rangle=\mathrm{reg}$. The algebra satisfied by both the $L_n$ and $\bar{L}_n$ is called the Virasoro algebra \cite{Virasoro}. To start with, we observe that except for the \textit{central} term containing the central charge $c$, the Virasoro algebra is exactly the same as the Witt algebra (c.f. \S \ref{sec:francescodeq2}). The reader can find a short discussion about the appearance of the central charge in appendix \ref{sec:apendicecentral}. Moreover, we note that we find two copies of the global conformal algebra $\frak{so}(3,1)\cong \frak{sl}(2,\mathbb{C})$ as the subalgebra generated by $L_{-1}, L_0$ and $L_{+1}$ and $\bar{L}_{-1}, \bar{L}_0$ and $\bar{L}_{+1}$ respectively. Notably, the central term affects only the structure of the local symmetry algebra given that for $n=0,\pm 1$ it's absent. Furthermore, the Witt generators of dilatations $(z,\bar{z})\to \lambda (z,\bar{z})$ in the $z-$plane, given by $\ell_0$ and $\bar{\ell}_0$ become the quantum generators $L_0$ and $\bar{L}_0$. As we said before, in radial quantization, time evolution is replaced with radial evolution, so that $L_0+\bar{L}_0$ will be proportional to the Hamiltonian.

\subsubsection{The Hilbert Space}
\label{sec:theHilbertSpace}
Having found a mechanism to realize the local conformal symmetry algebra through the action of mode operators acting within radially ordered correlation functions, we may now properly discuss the structure of the Hilbert space. Denoting the Hilbert space by $\mathcal{H}$ we take:
\begin{equation}
\label{ec:hilbertspace}
    \mathcal{H}=\bigoplus_{\mathcal{R},\bar{\mathcal{R}}} m_{\mathcal{R},\bar{\mathcal{R}}'}\mathcal{R}\otimes \bar{\mathcal{R}}'\,
\end{equation}
where $\mathcal{R}$ and $\bar{\mathcal{R}}$ are the representation spaces corresponding to different irreducible representations of the Virasoro algebra (related to the mode operators $L_n$ and $\bar{L}_n$ respectively), with corresponding multiplicities $m_{\mathcal{R},\bar{\mathcal{R}}'}$. Note that since the modes $L_n$ and $\bar{L}_m$ are decoupled (eq \eqref{ec:decoupledmodes}) we simply consider the tensor product of its representations. At first sight, equation \eqref{ec:hilbertspace} may look unnecessarily complex, but structuring the Hilbert space in terms of the representation spaces of the symmetry algebra of the system is, as we mentioned at the beginning of this section, a well known strategy. For instance, when studying the theory of angular momentum in standard quantum mechanics, we decompose our Hilbert space as the direct sum of representation spaces corresponding to different irreducible representations of the symmetry algebra, which in that case, is $\frak{su}(2)$.  

Continuing with the analogy of angular momentum, we construct the representation spaces $\mathcal{R}$ and $\bar{\mathcal{R}}$ by considering \textit{highest weight representations} \cite{Kac, Feigin, Kac2}. The construction for both the holomorphic and antiholomorphic sector is equivalent, so from now on we will focus on the holomorphic sector only. We choose a single generator $L_0$ to be diagonal in the representation space. Notably, since no pair of generators of the set $\{L_n\}$ commute, only one generator can be diagonalized at a time. We choose $L_0$ because it is the generator of dilatations, and as we said above, we can interpret it as the Hamiltonian of the system. We denote by $|h\rangle$ the highest weight state such that
\begin{equation}
    L_0|h\rangle = h |h\rangle\,.
\end{equation}
Since $[L_0,L_m]=-mL_m$ we have that:
\begin{equation}
   L_0 L_m|h\rangle = (h-m)L_m|h\rangle
\end{equation}
and so $L_m$ is a lowering operator for $m>0$. Similarly, $L_m$ is a raising operator for $m<0$. We adopt the condition 
\begin{equation}
    L_m|h\rangle = 0  \quad \text{for} \quad m>0\,,
\end{equation}
given that if we interpret $h$ to be the energy of the state (since $L_0$ is taken to be the Hamiltonian) we want the energy spectrum to be bounded from below\footnote{The choice $m>0$ is in fact arbitrary: we could as well take $m>r$ for any $r>0$, but this would not modify the analysis at all.}. On the other hand, we note that acting with $L_m$ produces a new eigenstate of $L_0$ with eigenvalue $h+m$. Hence, we may construct a basis of the representation space by applying the raising operators in all possible ways:
\begin{equation}
\label{ec:descendantstate}
    L_{-k_1}L_{-k_2}\dots L_{-k_n}|h\rangle \quad \text{with} \quad 1\leq k_1 \leq \dots \leq k_n\,.
\end{equation}
The state defined in equation \eqref{ec:descendantstate} is called a \textit{descendant state} and again, it is an eigenstate of $L_0$ with eigenvalue:
\begin{equation}
    h'=h+k_1+k_2+\dots+k_n\,.
\end{equation}
The representation space spanned by the highest weight state $|h\rangle$ and its descendants is called a \textit{Verma module} and we denote it by $V(c,h)$ where $c$ is the central charge and $h$ is the highest weight. The Hilbert space will then be decomposed as
\begin{equation}
    \mathcal{H}=\bigoplus_{(c,h),(c,\bar{h})}m_{(c,h),(c,\bar{h})}V(c,h)\otimes \bar{V}(c,\bar{h})
\end{equation}
where again, $m_{(c,h),(c,\bar{h})}$ is the multiplicity.

\subsubsection{Asymptotic States}
\label{sec:asymptoticstates}

Now that we have outlined the structure of the Hilbert space, we make contact with the field theory description from the preceding sections. To do so, we assume the existence of a vacuum state $|0\rangle$, which we take to be invariant under the global conformal group. This means that it should be anihilated by $L_{-1},L_0$ and $L_{+1}$ and their holomorphic counterparts, which sets the ground state energy to zero. This, in turn can be recovered from the condition that $T(z)|0\rangle$ and $\bar{T}(\bar{z})|0\rangle$ are well defined as $z,\bar{z}\to 0$, which implies:
\begin{equation}
\begin{array}{c}
    L_n |0\rangle=0  \\
    \bar{L}_n|0\rangle = 0 
\end{array}
\quad \text{for}\quad n\geq -1\,.
\end{equation}
Next, we define an asymptotic state as we usually do in QFT; by acting on the vacuum with a field in the limit in which $x^0\to -\infty$, where we assume it is free of interactions. Notably, when mapping the cylinder to the complex plane $x^0=-\infty$ is mapped to $z=0$ (c.f. equation \eqref{ec:conformalmap}) so that the asymptotic state is defined as
\begin{equation}
\label{ec:asymptoticstate}
    |\phi_{in}\rangle=\lim_{z,\bar{z}\to 0}\phi(z,\bar{z})|0\rangle\,.
\end{equation}
A small digression is in order. When defining a Hilbert space structure in a quantum field theory, the first step is to foliate our spacetime with respect to a certain coordinate. In regular QFTs, we usually choose the time coordinate so that we define a Hilbert space in each \textit{spatial slice} of fixed time. Evolution of states is then dictated by the time evolution operator, which is obtained by exponentiating the Hamiltonian $H\sim \partial_{x^0}$. When mapping the cylinder to the complex plane (c.f. \eqref{ec:conformalmap}) time evolution is replaced with radial evolution, so that in this context, the natural procedure is to define the Hilbert space on circles of constant radius over the complex plane. Evolution is still dictated by the Hamiltonian, which under the conformal map \eqref{ec:conformalmap} becomes the dilatation operator. Coming back to our definition of the asymptotic state in equation \eqref{ec:asymptoticstate}, we note that the surface of constant radius where the state is defined is very special: It is given by the single point $z=0$! Indeed, the existence of this spatial surface given by a unique point is a direct consequence of the conformal map between the cylinder and the complex plane, which maps the spatial slice $x^{0}=-\infty$ to $z=0$. This is a very distinctive feature of conformal field theories because it is the key element behind the \textit{state-operator correspondence}, which asserts that there exist a one to one correspondence between states of the Hilbert space and local operators. We will not provide a proof of this statement, which can be easily found in \cite{tong, Polchinski,SimonsDuffin}, instead we use it in appendix \ref{sec:apendiceOPE} to argue why the OPE is in fact an exact statement in CFTs.

Having introduced asymptotic states, we define the operation of Hermitian conjugation. In Minkowski space, Hermitian conjugation does not affect the time coordinate $t$, however in Euclidean space where the Euclidean time is given by $x^0 = it$, for $t$ to remain unchanged we require $x^0 \to -x^0$. In radial quantization, this corresponds to the mapping $z\mapsto 1/z^*$. Hence, we define Hermitian conjugation as:
\begin{equation}
    \phi(z,\bar{z})^{\dagger}=\bar{z}^{-2h}z^{-2\bar{h}}\phi(1/\bar{z},1/z)
\end{equation}
where $\phi(z,\bar{z})$ is by assumption a quasi-primary field of conformal dimensions $(h,\bar{h})$. The $z$ and $\bar{z}$ prefactors are justified by demanding that the asymptotic \textit{out} state:
\begin{equation}
    \langle \phi_{out}|=|\phi_{in}\rangle^\dagger
\end{equation}
has a well defined inner product with $|\phi_{in}\rangle$. Explicitly, we have
\begin{equation}
\label{ec:hermitianproduct}
\begin{split}
    \langle \phi_{out}|\phi_{in}\rangle 
    &= \lim_{z,\bar{z},w,\bar{w}\to 0}\langle 0 |\phi(z,\bar{z})^\dagger\phi(w,\bar{w})|0\rangle    \\
    &= \lim_{z,\bar{z},w,\bar{w}\to 0}\bar{z}^{-2h}z^{-2\bar{h}}\langle 0 |\phi(1/\bar{z},1/z)\phi(w,\bar{w})|0\rangle\\
    &= \lim_{\xi,\bar{\xi}\to \infty} \bar{\xi}^{2h}\xi^{2\bar{h}}\langle 0|\phi(\bar{\xi},\xi)\phi(0,0)|0\rangle \,.
\end{split}
\end{equation}
As a matter of fact, if we replace the general (fixed) form of the two point function given in equation \eqref{ec:twopointfunctionalmostorth}, we note that the $\xi$ and $\bar{\xi}$ factors of equation \eqref{ec:hermitianproduct} cancel with those appearing in the denominator of \eqref{ec:twopointfunctionalmostorth} providing a well defined $\xi,\bar{\xi}\to \infty$ limit and thus, a well defined inner product. 

Next, we relate our fields with states from the abstract Hilbert space introduced in \S \ref{sec:theHilbertSpace}. Let $\phi(z,\bar{z})$ be a primary field of conformal dimensions $(h,\bar{h})$ and let $|\phi_{in}\rangle$ be the asymptotic state obtained by inserting $\phi$ in the origin as in \eqref{ec:asymptoticstate}. Then  $|\phi_{in}\rangle$ is a highest weight state $|h,\bar{h}\rangle$. To see this, we start by computing the commutator:
\begin{equation}
\label{ec:comutatorprimaryfield}
\begin{split}
    [L_n,\phi(w,\bar{w})]
    &=\frac{1}{2\pi i} \oint_w dz \, z^{n+1} T(z)\phi(w,\bar{w})\\
    &=\frac{1}{2\pi i}\oint_w dz \, z^{n+1}\left[\frac{h\phi(w,\bar{w})}{(z-w)^2}+\frac{\partial \phi(w,\bar{w})}{z-w}+\mathrm{reg}\right]\\
    &=h(n+1)w^n\phi(w,\bar{w})+w^{n+1}\partial \phi(w,\bar{w}) \quad (n\geq -1)\,,
\end{split}
\end{equation}
were we have used the OPE of the stress tensor with a primary field given in equation \eqref{ec:primaryfieldope1}. There is a completely analogous expression for the $\bar{L}_n$. In particular, for $n=0$ we have that:
\begin{equation}
\label{ec:elecerocommutator}
\begin{split}
    L_0 \phi(0,0)|0\rangle
    &= ([L_0,\phi(0,0)]+\phi(0,0)L_0)|0\rangle\\
    &=h \phi(0,0)|0\rangle \\
    \bar{L}_0\phi(0,0)|0\rangle &= \bar{h}\phi(0,0)|0\rangle
\end{split}
\end{equation}
so that $|\phi_{in}\rangle=\phi(0,0)|0\rangle := |h,\bar{h}\rangle$ is an eigenstate of both $L_0$ and $\bar{L}_0$ with eigenvalues $h$ and $\bar{h}$ respectively. Furthermore, using the commutator \eqref{ec:comutatorprimaryfield} once again, we verify that:
\begin{equation}
\begin{array}{c}
    L_n |h,\bar{h}\rangle=0  \\
    \bar{L}_n|h,\bar{h}\rangle = 0 
\end{array}
\quad \text{if}\quad n>0
\end{equation}
which confirms that $|h,\bar{h}\rangle$ is in fact a highest weight state. 

Thus, we may now fully appreciate the importance of primary fields, which can be uniquely identified, when inserted at the origin, with the highest weight states of a Virasoro representation. In particular, for every primary field appearing in a certain theory, we will have a corresponding Verma module in the decomposition of the Hilbert space. Morevoer, as we will discuss in the next section, every non primary operator appearing in the theory will be uniquely identified with a descendant state of the appropiate Verma module. The systematic analysis of fields from this algebraic perspective, will be the main course of \S \ref{sec:chiralalebra}. In particular, we will see how the infinite dimensionality of the Virasoro algebra will impose sufficiently strong constraints to enable us to obtain results and perform computations that would not be possible otherwise: Ultimately, this is the reason why we did all of this construction.


\subsection{Operator Algebra}
\label{sec:chiralalebra}

In \S \ref{sec:theHilbertSpace} we wrote down our Hilbert space as a direct sum of tensor products of Verma modules. In this section we will exploit this algebraic structure to extract general results regarding two dimensional conformal field theories \cite{Belavin, Zamolodchikov2}. We start by defining the concept of \textit{descendant fields} which will be the fields related to the previously introduced descendant states. In particular, every field appearing in a theory will be either a primary or a descendant. Next we will show that the correlation functions of descendant fields can always be expressed in terms of correlation functions of the corresponding primaries, so that the computation of every correlation function in the theory reduces to the calculation of correlation functions between  primary fields. In particular, by computing the OPE of the primary fields appearing within the correlation functions, we may reduce the number of points down to two point functions, which are completely fixed by conformal symmetry and thus, the theory is completely solved. Notably, the complete OPE between all primary fields is not fixed by conformal symmetry and some dynamical input is necessary to compute the OPE coefficients. We will discuss how in some very special cases, this procedure renders a completely solvable theory.

\subsubsection{Descendant Fields}
\label{sec:descendantfields}

In \S \ref{sec:theHilbertSpace} we defined asymptotic states by acting with primary fields inserted in the origin over the vacuum $\phi(0,0)|0\rangle$, which we then verified were highest weight states. Thus, for each primary field we have an infinite number of descendant states obtained by acting arbitrarily with the mode operators $L_n$ and $\bar{L}_n$ respectively. In other words, for each primary field we have an entire Verma module, which by construction, is invariant under conformal transformations. 

Descendant states may also be interpreted as the result of the application of a \textit{descendant field} over the vacuum. For example, the descendant state $L_{-n}|h\rangle$ can be expressed as:
\begin{equation}
\label{ec:descendantfield}
    L_{-n}|h\rangle=L_{-n} \phi(0)|0\rangle = \frac{1}{2\pi i}\oint dz \, z^{1-n}\,T(z)\phi(0)|0\rangle
\end{equation}
where we have dropped the antiholomorphic part to simplify notation. Looking at the above equation, we may define the descendant field corresponding to the descendent state $L_{-n}|h\rangle$ as:
\begin{equation}
    \phi^{(-n)}(w):=\frac{1}{2\pi i}\oint_w dz\,  \frac{1}{(z-w)^{n-1}}\,T(z)\phi(w)\quad \text{so that}\quad \phi^{(-n)}(0)|0\rangle=L_{-n}|h\rangle\,.
\end{equation}
In particular, since $\phi$ is a primary field, we may use the OPE \eqref{ec:primaryfieldope1} to compute:
\begin{equation}
    \phi^{(0)}(w)=h\phi(w) \quad \text{and}\quad \phi^{(-1)}(w)=\partial \phi(w)\,.
\end{equation}
The fundamental property of descendant fields is that their correlation functions can be derived from the corresponding primaries. To see this, let us consider the correlator $\langle \phi^{(-n)}(w)X\rangle$ where $X=\phi_1(w_1)\dots \phi_n(w_n)$ is a product of primary fields of dimensions $h_i$. Then, 
\begin{equation}
\label{ec:descendant1}
\begin{split}
     \langle \phi^{(-n)}(w)X\rangle
     &= \frac{1}{2\pi i}\oint_w dz\, (z-w)^{1-n}\langle T(z)\phi(w) X\rangle \\
     &= -\frac{1}{2\pi i}\oint_{\{w_i\}}dz \, (z-w)^{1-n}\sum_{i}\left[\frac{1}{z-w_i}\partial_{w_i}\langle \phi(w)X\rangle +\frac{h_i}{(z-w_i)^2}\langle \phi(w)X\rangle\right]\\
     &=\sum_{i}\left[\frac{(n-1)h_i}{(w_i-w)^n} -\frac{1}{(w_i-w)^{n-1}}\partial_{w_i}\right]\langle \phi(w)X\rangle \\
\end{split}
\end{equation}
where in the first line we used the definition of the descendant field with the contour enclosing only $w$ and excluding the $w_i$; in the second line we reversed the contour so that the $w_i$ are now inside the region of integration, and we used the $T\phi$ OPE given in equation \eqref{ec:primaryfieldope1}. In the third line, we used the residue theorem. If we further define the operator
\begin{equation}
    \mathcal{D}_{-n}:=\sum_{i}\left[\frac{(n-1)h_i}{(w_i-w)^n} -\frac{1}{(w_i-w)^{n-1}}\partial_{w_i}\right]\,,
\end{equation}
equation \eqref{ec:descendant1} becomes:
\begin{equation}
    \langle \phi^{(-n)}(w)X\rangle=\mathcal{D}_{-n}\langle \phi(w)X\rangle
\end{equation}
so that the correlator of $X$ with a descendant field can be obtained by acting with a differential operator on the correlator of $X$ with the original primary field. 

Notably, the descendant field defined in equation \eqref{ec:descendantfield} is not the most general possibility given that we may as well consider an arbitrary descendant state $L_{-k_1}\dots L_{-k_n}|h\rangle$. The corresponding descendant field is defined recursively:
\begin{equation}
       \phi^{(-k_1,\dots,-k_n)}(w)
       =\frac{1}{2\pi i}\oint_w dz\, (z-w)^{1-k_1} T(z)\phi^{(-k_2,\dots,-k_n)}(w)\,. 
\end{equation}
The set comprising a primary field and all of it descendant fields is called a \textit{conformal family} and it is sometimes denoted by $[\phi]$. In particular, it can be shown that the correlator of an arbitrary descendant field with a product of primary fields $X$ can be expressed in terms of the corresponding primary:
\begin{equation}
    \langle \phi^{(-k_1,\dots,-k_n)}(w)X\rangle = \mathcal{D}_{-k_1}\dots \mathcal{D}_{-k_n}\langle \phi(w)X\rangle\,.
\end{equation}
We may also consider correlators with more than one descendant field but the result is exactly the same: These can always be reduced to correlators including solely the corresponding primaries. Hence, given that every field that is not primary is necessarily a descendant (after the identification with the corresponding descendant state), we conclude that any correlation function in the theory can be expressed in terms of correlation functions of primary fields. 

\subsubsection{The Operator Algebra}

We have seen that the correlation function of any product of fields appearing in a two dimensional conformal field theory can be reduced to a correlator including only primary fields. Hence, we may restrict our study to correlation functions of primary fields only. Moreover, with the OPE at hand, we know that we may reduce $n-$point functions to $(n-1)-$point functions and if we do this recursively, we may reduce any $n-$point function into a two point function\footnote{This procedure of iteratively reducing the number of points in a correlation function strongly relies on the exactness of the OPE. This is why we can't proceed in this way in field theories without conformal invariance.}, so let us start with the latter.  In \S \ref{sec:twopointfunction} we saw that the two point function is completely fixed by conformal symmetry to be:
\begin{equation}
    \langle \phi_{i}(z_i,\bar{z}_i)\phi_{j}(z_j,\bar{z}_j) \rangle =\frac{C_{ij}}{(z_i-z_j)^{2h}(\bar{z}_i-\bar{z}_j)^{2\bar{h}}} \quad \text{if} \quad 
    \begin{cases}
    \begin{array}{c}
         h_1=h_2=h  \\
         \bar{h}_1=\bar{h}_2=\bar{h} 
    \end{array}
    \end{cases} \,,
\end{equation}
where as we said before, the two point function vanishes if the fields have different conformal dimension. In particular, the coefficients $C_{ij}$ are symmetric, and thus, with a suitable normalization, we may choose the primary fields such that $C_{ij}=\delta_{ij}$. In particular, for this choice of normalization, we have that conformal families corresponding to different primary fields are orthogonal in the sense of the two point function. As a matter of fact, this is a consequence of the orthogonality of Verma modules corresponding to different heighest weight representations: By a suitable conformal transformation we can always bring the points $z_i$ and $z_j$ appearing in the correlator to $z_i=\infty$ and $z_j=0$ respectively so that the fields are then asymptotic and the two point function becomes an inner product in the Hilbert space:
\begin{equation}
    \lim_{z_i,\bar{z}_i\to \infty}z_i^{2h}\bar{z}_i^{2\bar{h}}\langle \phi_i(z_i,\bar{z}_i)\phi_j (0,0)\rangle = \langle h_i|h_j\rangle \langle \bar{h}_i|\bar{h}_j\rangle =\delta_{ij}\,.
\end{equation}
With the two point function properly normalized, we may now proceed to study OPE's between primary fields. We know that in general, an OPE is given by a sum over all of the operators appearing in the theory. Given that every operator is either a primary field or a descendant, we may write a general OPE between two primary fields as:
\begin{equation}
\label{ec:OPEprimaryfields}
    \phi_i(z,\bar{z})\phi_j(0,0)=\sum_{p}\sum_{\{k,\bar{k}\}}C_{pij}^{\{k,\bar{k}\}}z^{h_k-h_i-h_j+K}\bar{z}^{\bar{h}_k-\bar{h}_i-\bar{h}_j+\bar{K}}\phi^{\{k,\bar{k}\}}_p(0,0)\,,
\end{equation}
where $p$ runs over the set of primary fields and $\{k,\bar{k}\}$ represents the collection of indices $k_i$ and $\bar{k}_i$ of the corresponding descendant field. The powers of $z$ and $\bar{z}$ appearing in the expansion are fixed by scale invariance, where we have defined $K=\sum_i k_i$ and $\bar{K}=\sum_i \bar{k}_i$. Next, we recall from \S \ref{sec:twopointfunction} that the three point function is also fixed by conformal symmetry up to an unknown coefficient:
\begin{equation}
\label{ec:threepointfunction}
\left\langle\phi_{i}\left(z_{i}\right) \phi_{j}\left(z_{j}\right) \phi_{k}\left(z_{k}\right)\right\rangle
=C_{ijk} \frac{1}{z_{ij}^{h_{i}+h_{j}-h_{k}}  z_{jk}^{h_{j}+h_{k}-h_{i}} z_{ik}^{h_{k}+h_{i}-h_{j}}}\cdot
\frac{1}{\bar{z}_{ij}^{\bar{h}_{i}+\bar{h}_{j}-\bar{h}_{k}} \bar{z}_{jk}^{\bar{h}_{j}+\bar{h}_{k}-\bar{h}_{i}} \bar{z}_{ik}^{\bar{h}_{k}+\bar{h}_{i}-\bar{h}_{j}}}\,.
\end{equation}
Hence, if we take the correlator of \eqref{ec:OPEprimaryfields} with a third primary field $\phi_r(z_r,\bar{z}_r)$ of dimensions $(h_r,\bar{h}_r)$ and consider the asymptotic state defined taking $z_r,\bar{z}_r \to \infty$ we have:
\begin{equation}
\begin{split}
    \langle \phi_r|\phi_i(z_i,\bar{z}_i)|\phi_j\rangle 
    &= \lim_{z_r,\bar{z}_r\to \infty}z_r^{2h_r}\bar{z}_r^{2\bar{h}r}\langle \phi_r(z_r,\bar{z}_r)\phi_i(z_i,\bar{z}_i)\phi_j(0,0)\rangle  \\
    & = \frac{C_{rij}}{z_i^{h_i+h_j-h_r}\bar{z}_i^{\bar{h}_i+\bar{h}_j-\bar{h}_r}}\,,
\end{split}
\end{equation}
where the second line follows by replacing \eqref{ec:threepointfunction} in the first line and taking $z_r,\bar{z}_r\to \infty$, noting that the factors $z_r^{2h_r}$ and $\bar{z}_r^{2\bar{h}_r}$ cancel with the $z_{ir}$, $z_{jr}$ and $\bar{z}_{ir}$, $\bar{z}_{jr}$ respectively, appearing in the denominator. On the RHS of \eqref{ec:OPEprimaryfields}, when taking the correlator with $\phi_r(z_r,\bar{z}_r)$, we have that the only non vanishing term, given the orthogonality of primary fields and descendants, will be given by $p\{k,\bar{k}\}=r\{0,0\}$, so we obtain the relation:
\begin{equation}
   C_{rij}= C_{rij}^{\{0,0\}}\,.
\end{equation}
More generally, may write:
\begin{equation}
    C_{rij}^{\{k,\bar{k}\}}=C_{rij}\beta_{rij}^{\{k\}}\bar{\beta}_{rij}^{\{\bar{k}\}}
\end{equation}
which implies that a descendant field can have a non vanishing correlator with another field only if the corresponding primary is correlated with this field. The coefficients $\beta_{rij}^{\{k\}}$ and $\bar{\beta}_{rij}^{\{\bar{k}\}}$ are determined as functions of the conformal dimensions $(h,\bar{h})$ and the central charge $c$, by requiring that both sides of equation \eqref{ec:OPEprimaryfields} transform identically under conformal transformations.

We see that the complete information to specify a two dimensional conformal field theory is provided by the conformal weights $(h_i, \bar{h}_i)$ of the Virasoro
highest weight states corresponding to the primary fields, the central charge $c$ and OPE coefficients $C_{ijk}$. Everything else follows from the values of these parameters, which themselves cannot be determined solely on the basis of conformal symmetry.

\subsubsection{Crossing Symmetry}

\label{sec:crossingsymmetry}

\begin{figure}[h]
    \centering
    \includegraphics{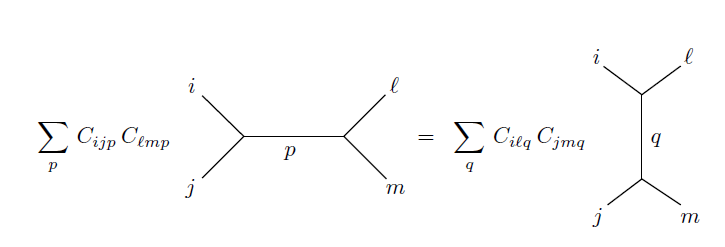}
    \caption{Schematic figure of the crossing symmetry constraint.}
    \label{fig:crossingsymmetry}
\end{figure}

To determine the unspecified $C_{ijk}$ and the conformal weights, we need to apply some dynamical principle to obtain additional information. To see how this works, we consider the four point function
\begin{equation}
\left\langle\phi_{1}\left(z_{1}, \bar{z}_{1}\right) \phi_{2}\left(z_{2}, \bar{z}_{2}\right) \phi_{3}\left(z_{3}, \bar{z}_{3}\right) \phi_{4}\left(z_{4}, \bar{z}_{4}\right)\right\rangle\,.
\end{equation}
Since the order in which fields appear within the correlation function doesn't matter, we have different ways of computing the double OPE, which correspond to the different choices of the fields that intervene in the single OPE. For example, we can compute the four point function as the double OPE
\begin{equation}
\label{ec:fourpoint1}
\left\langle\left[\phi_{1}\left(z_{1}, \bar{z}_{1}\right) \phi_{2}\left(z_{2}, \bar{z}_{2}\right)\right]\left[ \phi_{3}\left(z_{3}, \bar{z}_{3}\right) \phi_{4}\left(z_{4}, \bar{z}_{4}\right)\right]\right\rangle
\end{equation}
where the square brackets indicate that the OPE is taken between the fields therein. But we could as well, consider:
\begin{equation}
\label{ec:fourpoint2}
\left\langle\left[\phi_{1}\left(z_{1}, \bar{z}_{1}\right)\phi_{3}\left(z_{3}, \bar{z}_{3}\right) \right]\left[\phi_{2}\left(z_{2}, \bar{z}_{2}\right)  \phi_{4}\left(z_{4}, \bar{z}_{4}\right)\right]\right\rangle\,.
\end{equation}
If we want our operator algebra to be associative, then equations \eqref{ec:fourpoint1} and \eqref{ec:fourpoint2} should coincide. Notably, in $d=2$ the associativity of the four point function implies the associativity  of any higher $n-$point functions so it is a necessary and sufficient condition for the operator algebra to be associative\footnote{This is not necessarily true in $d>2$, c.f. \cite{Rychkov2}}. This requirement is a consistency condition, known as the crossing symmetry of the four point function, which provides a system of algebraic equations for the three point function coefficients $C_{ijk}$ as shown in figure \ref{fig:crossingsymmetry}. 

The procedure of solving the relations of figure \ref{fig:crossingsymmetry} is known as the \textit{conformal bootstrap}. This program is usually very difficult to implement because of the possibly infinite number of unknowns that intervene. In two dimensions the situation is partially simpler, because we need only to consider primary fields, vastly reducing the number of independent quantities. Nonetheless, it still remains the possibility that there is an infinite number of primary fields and thus making any progress remains to be a very difficult task. 

Notably, there are some specific situations in which the conformal bootstrap equations are known to be exactly solvable. These are the cases of the minimal models, in which the number of conformal primaries is finite \cite{Cardy2, Belavin, Friedan3, Bershadsky, Fateev2,Difrancesco1, Difrancesco2}. A slightly more general example is that of rational or meromorphic conformal field theories, such as the WZW models in which the possibly infinite irreducible representations of the Virasoro algebra can be reorganized in a finite number of blocks of a certain extended symmetry. This is the idea underlying extended chiral algebras, which we introduce in the following section.

\section{Chiral Algebras}
\label{sec:chiralalgebra}

In the previous section we discussed thoroughly the basics of two dimensional conformal field theories and outlined its implications. We found that the local symmetry algebra of conformal transformations could be realized at the quantum level in terms of the modes of the holomorphic and anti-holomorphic components of a stress energy tensor, which as we said in \S \ref{sec:OPE}, are quasi-primary fields of conformal dimensions $(2,0)$ and $(0,2)$ respectively. More generally, we will refer to quasi-primary fields having conformal dimensions $(h,0)$ as \textit{chiral fields} and similarly, for fields with conformal dimensions $(0,\bar{h})$ as \textit{anti-chiral} fields. As a matter of fact, all of the algebraic structure regarding the Virasoro algebra discussed in the previous section, is a direct consequence of the existence of a chiral and anti-chiral field, given by the components of the stress tensor. 

Generally speaking, chiral symmetry arises as a consequence of the existence of a local operator $\mathcal{O}(z,\bar{z})$ satisfying a meromorphicity condition of the form:
\begin{equation}
    \partial_{\bar{z}}\mathcal{O}(z,\bar{z}) \quad \Longrightarrow \quad \mathcal{O}(z,\bar{z}):=\mathcal{O}(z)\,.
\end{equation}
Under these assumptions, such an operator will transform in the trivial representation of the anti-holomorphic part of the symmetry algebra, in other words, it is a chiral operator. Meromorphicity implies the existence of infinitely many conserved charges (and their associated Ward identities) given by the modes of the operator, which are obtained by integrating the meromorphic operator against an arbitrary monomial in $z$,
\begin{equation}
\label{ec:modesafterintegration}
    \mathcal{O}_n:=\oint \frac{dz}{2\pi i}\, z^{n+h-1}\mathcal{O}(z)\,.
\end{equation}
The operator product expansion of two meromorphic operators contains only meromorphic operators, and the singular terms determine the commutation relations among the associated charges. This is the power of meromorphy in two dimensions: an infinite dimensional symmetry algebra organizes the space of local operators into much larger representations, and the associated Ward identities strongly constrain the correlation functions of the theory. 

With this in mind, we define a \textit{chiral algebra} to be the infinite dimensional symmetry algebra associated to the modes of a chiral operator. As we mentioned in the beginning of this article, chiral algebras are defined in slightly different ways depending on the literature. For instance, in the context of vertex operator algebras, even axiomatic definitions can be found \cite{Bouwknegt, Goddard3}. It is not the aim of this paper to go into the theory of vertex operator algebras so we will content ourselves with the definition we gave above. In particular, we observe that the modes of each of the components of the stress tensor generate a chiral algebra, namely, the Virasoro algebra which is indeed our first and most important example. 

The idea of structuring the Hilbert space in terms of a symmetry algebra realized through the modes of a chiral field, raises the question about the existence of other chiral fields other than the stress energy tensor, that could lead to additional or extended symmetries, or making use of the terminology, extended chiral algebras. This was first proposed in the pioneering paper \cite{Zamolodchikov1}, in which the existence of additional infinite dimensional symmetry in two dimensional conformal field theories was introduced. The power of these extended chiral algebras relies in the fact that the constraints imposed by this additional symmetry, together with those imposed by the Virasoro algebra, are in many cases, enough to completely solve a theory.

Since these extended chiral algebras were introduced, different approaches have been employed in their description, but more importantly, an impressive amount of work has been done both from the physics and the mathematics side \cite{Rastelli1,Moore2, Moody, Goddard2}. The objective of this paper is not to do a review on extended chiral algebras (we already have a great one \cite{Bouwknegt}), but instead to provide the necessary background to engage with the topic in a serious manner. Hence, we will restrict ourselves to present one of the paramount examples of a system with an extended chiral algebra, known as the Wess-Zumino-Witten (WZW) model. This will serve to illustrate both the mechanism in which an extended symmetry may appear, and most importantly, how the extension of the chiral algebra provides additional constraints which allow the exact solvability of the model. 


\subsection{Wess-Zumino-Witten Models}
\label{sec:WZW}

Wess-Zumino-Witten (WZW) models \cite{Zamolodchikov1, Witten, Wess, Knizhnik, Gepner, Verlinde3} have played a fundamental role in modern high energy physics, due to its applications in both string theory and in the study of conformal field theories with extended symmetries. We will introduce the topic in a superficial manner, given that we are only interested in the chiral algebra structure of the model. There is an endless number of sources discussing WZW models (see \cite{Frohlich, blumenhagen, Francesco, Eberhardt2} and references therein). 

\subsubsection{Non Linear Sigma Models}

The first step in order to introduce WZW models is to consider a \textit{non linear sigma model} describing bosonic degrees of freedom whose target space is given by a group manifold $G$. The action is given by:
\begin{equation}
\label{ec:nonlinearsigmamodel}
    S_0=\frac{1}{4a^2}\int d^2x \, \mathrm{Tr}(\partial^\mu g^{-1}\partial_\mu g)
\end{equation}
where $a^2$ is a positive dimensionless coupling and $g(x)$ is a matrix bosonic field taking values on the group manifold $G$. We can obtain the equations of motion by studying the variation of the action under the variation $g\to g+\delta g$, which leads to the conservation of the currents
\begin{equation}
    J_\mu = g^{-1}\partial_\mu g
\end{equation}
which in complex coordinates may be written as $\tilde{J}_z=g^{-1}\partial_z g$ and $\tilde{J}_{\bar{z}}=g^{-1}\partial_{\bar{z}}g$ so that:
\begin{equation}
    \partial_z \tilde{J}_z +\partial_{\bar{z}}\tilde{J}_{\bar{z}}=0\,.
\end{equation}
Notably, these currents are not conserved separately and thus, we can't speak about holomorphic and anti-holomorphic quantities. 

\subsubsection{The WZW Term}

In order to obtain holomorphic and anti-holomorphic currents, we require the addition of a WZW term \cite{Wess} to the action, given by:
\begin{equation}
    \Gamma = \frac{-i}{24\pi}\int_B d^3 y \, \epsilon_{\alpha \beta \gamma}\mathrm{Tr}(\tilde{g}^{-1}\partial^{\alpha}\tilde{g}\tilde{g}^{-1}\partial^\beta\tilde{g}\tilde{g}^{-1}\partial^\gamma\tilde{g}) \,.
\end{equation}
The integral is defined over a three dimensional manifold $B$, whose boundary is the compactification of our original two dimensional space. Thus, $\tilde{g}$ is the extension of the field $g(x)$ to $B$. Notably, the extension of $g(x)$ to the three dimensional manifold is not unique, and it can be shown that in order to deal with these ambiguity, the structure of the total action should be:
\begin{equation}
\label{ec:wzwaction}
    S^{WZW}=S_0+k\Gamma
\end{equation}
with $k$ an integer\footnote{We are intentionally leaving all of the details aside in order to not deviate from the main point. The complete development can be found in \cite{Francesco, Novikov}.}. In particular, taking $a^2=4\pi/k$ in \eqref{ec:nonlinearsigmamodel} and computing the equations of motion, we find the separately conserved holomorphic and anti-holomorphic currents:
\begin{equation}
    J_z = \partial_z g g^{-1} \quad \text{and} \quad J_{\bar{z}}=g^{-1}\partial_{\bar{z}}g\,.
\end{equation}
The separate conservation of $J_z$ and $J_{\bar{z}}$ implies the invariance of the action under:
\begin{equation}
\label{ec:WZWtransformation}
    g(z,\bar{z})\to \Omega(z)g(z,\bar{z})\bar{\Omega}^{-1}(\bar{z})
\end{equation}
where $\Omega(z)$ is a holomorphic map taking values in $G$ and similarly with $\bar{\Omega}(z)$. In other words, $J_z$ and $J_{\bar{z}}$ are the conserved currents associated to the transformation \eqref{ec:WZWtransformation}. Notably, the original action $S_0$ is globally $G\times G$ invariant, and the WZW term enhances the symmetry to local $G(z)\times G(\bar{z})$ invariance. 


\subsubsection{Current Algebra}

Having found an action with a local symmetry, we analyze the structure of the underlying Lie algebra. We start by rescaling our currents as:
\begin{equation}
\label{ec:wzwcurrents}
\begin{split}
    J(z)&:=-kJ_z(z)=-k\partial_z g g^{-1}  \\
    \bar{J}(\bar{z})
    &:= kJ_{\bar{z}}(\bar{z})=kg^{-1}\partial_{\bar{z}} g \,,
\end{split}
\end{equation}
where $k$ is the integer appearing in  \eqref{ec:wzwaction}. Morevoer, the currents coincide with the pullback of the (left and right) Maurer-Cartan form to $\mathbb{C}^2$ under the map defined by $g:\mathbb{C}^2\to G$ \cite{Eberhardt2,Nakahara}, and thus, we may uniquely identify them with elements of the corresponding Lie algebra $\frak{g}=\mathrm{Lie}(G)$. Therefore, we may express them in terms of their components in a basis of $\frak{g}$ as:
\begin{equation}
\label{ec:currentcomponents}
    J(z)=\sum_a J_{a}(z)t^a \quad \text{and}\quad \bar{J}(\bar{z})=\sum_a \bar{J}_a(\bar{z})t^a\,.
\end{equation}
As usual, we focus on the holomorphic current given that the anti-holomorphic counterpart is completely analogous. Proceeding in the same way than in section \ref{sec:2dCFT} we may compute the associated Ward identities by considering an infinitesimal version of the transformation \eqref{ec:WZWtransformation}, and find the OPE of the components of the current to be \cite{Francesco}:
\begin{equation}
\label{ec:currentalgebraOPE}
    J^{a}(z)J^{b}(z)=\frac{k \delta^{ab}}{(z-w)^2}+\sum_c i f_{abc}\frac{J^c(w)}{(z-w)}+\mathrm{reg}
\end{equation}
where $f_{abc}$ are the structure constants of $\frak{g}$ and with a completely analogous expression for the anti-holomorphic sector. The OPE given in \eqref{ec:currentalgebraOPE} defines a \textit{current algebra}. As we did with the stress tensor, we expand the components of the currents as:
\begin{equation}
\label{ec:currentsme}
    J^a(z)=\sum_{n\in \mathbb{Z}}z^{-n-1}J_n^a
\end{equation}
and compute the commutation relations between them as we did for the Virasoro algebra (c.f. \S \ref{sec:virasoroalgebra}):
\begin{equation}
\label{ec:kacmoodyalgebra}
    \left[J_n^a,J_m^b\right]= i\sum_c f_{abc}J^c_{n+m}+kn\delta^{ab}\delta_{n+m,0}\,.
\end{equation}
The commutation relations given in \eqref{ec:kacmoodyalgebra} define what is known in the literature as an affine Kac-Moody algebra of level $k$, usually denoted as $\hat{\frak{g}}_k$. There is an analogous expression for the anti-holomorphic counterpart, and since the currents are independent, we have
\begin{equation}
    \left[J_n^a,\bar{J}_m^b\right]=0\,.
\end{equation}
We observe, that in the same way than in two dimensional theories with conformal symmetry, the local symmetry algebra decouples into a tensor product of two copies of an infinite dimensional \textit{affine} Lie algebra. At this point the reader might wonder how is this related to conformal symmetry at all. Indeed, in the next section we'll see that the local symmetry algebra we have introduced, will prove to imply conformal invariance.  

\subsubsection{The Sugawara Construction}

In principle, we could verify that the theory defined by the action \eqref{ec:wzwaction} is (locally) conformally invariant by analyzing its properties under a conformal transformation. Although this would in principle satisfy our conformal thirst, we want to make local conformal symmetry manifest in a way that it is compatible with the local symmetry algebra corresponding to the Kac-Moody algebra introduced in the preceeding section. This is the essence of the \textit{Sugawara construction} \cite{Sugawara, Sommerfield}. 

The starting point is to construct a stress energy tensor in terms of the components of the currents $J^a(z)$. We take the ansatz:
\begin{equation}
    T(z)=\gamma \sum_a (J_a J^a)(z)
\end{equation}
where $(J_a J^a)(z)$ is the normal ordered product of operators defined by the formula:
\begin{equation}
\label{ec:normalorder}
    (J_aJ^a)(z)=\frac{1}{2\pi i}\oint_z \frac{dw}{w-z}J_a(w)J^a(z)
\end{equation}
and the constant factor $\gamma$ is an unspecified coefficient which we fix by demanding that $J^a(z)$ is a chiral primary field of dimension $(1,0)$. That is, that the OPE with the stress tensor has the form:
\begin{equation}
\label{ec:currentprimary}
    T(z)J^a(w)=\frac{J^a(w)}{(z-w)^2}+\frac{\partial J^a(w)}{(z-w)}+\mathrm{reg}\,.
\end{equation}
Again, we omit the details of the computation, but it can be shown \cite{Francesco, Dashen, Todorov} that \eqref{ec:currentprimary} requires 
\begin{equation}
\label{ec:Sugawarastresstensor}
    T(z)=\frac{1}{2(k+h^{\vee})}\sum_a (J_a J^a)(z)
\end{equation}
where $h^{\vee}$ is called the \textit{dual Coexeter number} of the affine Lie algebra $\hat{\frak{g}}_k$ defined by:
\begin{equation}
    f_{abc}f_{dbc}=2h^{\vee}\delta_{ab}\,.
\end{equation}
In particular, for the chosen value of $\gamma$ we have that the stress tensor's self OPE takes the form:
\begin{equation}
\label{ec:TTOPE2}
    T(z)T(w)=\frac{c/2}{(z-w)^4}+\frac{2 T(w)}{(z-w)^2}+\frac{\partial T(w)}{z-w} + \mathrm{reg} 
\end{equation}
with the central charge given by:
\begin{equation}
\label{ec:sugawaracentral}
    c=\frac{k\, \mathrm{dim}\, \frak{g}}{k+h^{\vee}}\,.
\end{equation}
Of course, everything can be equally done for the anti-holomorphic sector, and thus, equations \eqref{ec:TTOPE2} and \eqref{ec:sugawaracentral} imply the conformal invariance of the WZW model. We may further consider a mode expansion of the stress energy tensor
\begin{equation}
\label{ec:WZWstresstensor}
    T(z)=\sum_{n\in \mathbb{Z}}z^{-n-2}L_n \quad \text{with}\quad L_n=\frac{1}{2(k+h^{\vee})}\sum_m :J_m^a J_{n-m}^a:
\end{equation}
so that the complete affine Lie algebra and Virasoro algebra commutation relations are given by\footnote{This analysis has been assuming an orthonormal basis in terms of the Killing form. For a more general basis, the commutation relations of the currents would involve the Killing form as opposed to the Kronecker delta and this would then carry through into the stress tensor and the Virasoro modes.}:
\begin{equation}
\begin{split}
    [L_n,L_m]&=(n-m)L_{n+m}+\frac{c}{12}(n^3-n)\delta_{m+n,0}\\
    [L_n,J_m^a]&=-mJ^{a}_{n+m}\\
    [J_n^a,J_m^b]&=\sum_c i f_{abc}J^c_{n+m}+kn\delta_{ab}\delta_{n+m,0}\,.
\end{split}
\end{equation}
Notably, the zero mode of the current algebra commutes with the Virasoro generators (in particular with $L_0$) which reflects the built-in $\frak{g}-$invariance of the model. The full Lie algebra, however, does not commute with $L_0$ and in fact these will turn out to be the \textit{spectrum generating algebra} of the theory. 

\subsubsection{WZW Primary Fields}

Having obtained the commutation relations of the complete symmetry algebra, we proceed to analyze the algebraic structure of the WZW model. As in the purely conformal case, where we defined primary fields as those transforming in irreducible representations of the local conformal algebra, we may define WZW primary fields as those transforming covariantly with respect to a $G(z)\times G(\bar{z})$ transformation as in \eqref{ec:WZWtransformation}:
\begin{equation}
\label{ec:wzwtransformation2}
    \phi(z,\bar{z})\to \Omega(z)\phi(z,\bar{z})\bar{\Omega}^{-1}(\bar{z})\,.
\end{equation}
Let us express this transformation property in terms of an OPE. To do so, we consider an infinitesimal version of \eqref{ec:wzwtransformation2}, with
\begin{equation}
\begin{split}
    \Omega(z)
    &=1+\omega(z) \\
    &=1+\sum_a \omega^a(z)t^a
\end{split}
 \quad\quad\quad \quad
\begin{split}
 \bar{\Omega}(\bar{z})
    &=1+\bar{\omega}(\bar{z})\\
    &=1+\sum_a \bar{\omega}^a(\bar{z})t^a  
\end{split} 
\end{equation}
so that the variation of the field \eqref{ec:wzwtransformation2} becomes to first order in $\omega$:
\begin{equation}
\label{ec:wzwvariation}
\begin{split}
    \delta_{\omega,\bar{\omega}}\phi(z,\bar{z})
    &= \omega(z) \phi(z,\bar{z})-\phi(z,\bar{z})\bar{\omega}(\bar{z}) \,.
\end{split}
\end{equation}
On the other hand, in analogy to what we did in \S \ref{sec:conformalwardidentity}, we may compute the Ward Identity associated to a $G(z)\times G(\bar{z})$ transformation, which is given by \cite{Francesco}:
\begin{equation}
\label{ec:wzwvariation2}
    \delta_{\omega,\bar{\omega}}\langle X\rangle = -\frac{1}{2\pi i}\oint dz\, \omega^a(z)\langle J^a(z) X\rangle +\frac{1}{2\pi i}\oint d\bar{z}\,\bar{\omega}^a(\bar{z})\langle \bar{J}^a(\bar{z})X\rangle
\end{equation}
where $X=\phi_1(z_1,\bar{z}_1)\dots \phi_n(z_n,\bar{z}_n)$ is an arbitrary string of fields and $J^a(z)$ and $\bar{J}^a(\bar{z})$ are the components of the holomorphic and anti-holomorphic currents as defined in equation \eqref{ec:currentcomponents}. Hence, from equations \eqref{ec:wzwvariation} and \eqref{ec:wzwvariation2} we may compute, by means of the residue theorem, the OPE of a WZW primary field with the currents:
\begin{equation}
\label{ec:wzwprimaryope}
\begin{split}
    J^a(z)\phi_{\lambda,\mu}(w,\bar{w})&=-\frac{t_\lambda^a \, \phi_{\lambda,\mu}(w,\bar{w})}{z-w} +\mathrm{reg}\\
 \bar{J}^a(\bar{z})\phi_{\lambda,\mu}(w,\bar{w})&=\frac{\phi_{\lambda,\mu}(w,\bar{w})\, t_\mu^a }{\bar{z}-\bar{w}}+\mathrm{reg} \,.
\end{split}
\end{equation}
Note that we added the subscripts $\lambda,\mu$, which make reference to the representation of $G(z)\times G(\bar{z})$ under which the field is transforming: The holomorphic sector belongs to a representation with heighest weight $\lambda$, while the anti-holomorphic sector, to a representation with heighest weight $\mu$. In what follows, we restrict ourselves to the holomorphic sector. With the OPE at hand, we may define the state $|\phi_\lambda\rangle = \phi_\lambda(0)|0\rangle$ as in \S \ref{sec:asymptoticstates}, and analyze the action of the modes of the currents defined in \eqref{ec:currentsme}. With a completely analogous computation to the one in equation \eqref{ec:elecerocommutator}, we find that if $\phi_\lambda$ is a WZW primary field, 
\begin{equation}
\begin{cases}
\begin{split}
        J_0^a|\phi_\lambda\rangle &=-t_\lambda^a|\phi_\lambda\rangle \\
     J_n^a|\phi_\lambda \rangle &= 0 \quad \text{for}\quad n>0\,.
\end{split}
\end{cases}
\end{equation}
A remarkable property of WZW primaries is that they are also Virasoro primaries. To see this, from equation \eqref{ec:WZWstresstensor} we note that
\begin{equation}
    L_n|\phi_\lambda\rangle = 0 \quad \text{for}\quad n>0
\end{equation}
given that $:J^a_n J^a_{m-n}:$ has by definition the $J^a$ factor with the largest subscript to the right which anihilates $|\phi\rangle$ for $n>0$. On the other hand, 
\begin{equation}
\begin{split}
    L_0|\phi_\lambda \rangle 
    &= \frac{1}{2(k+h^{\vee})}\sum_a J_0^aJ_0^a|\phi_\lambda\rangle  \\
    &=\frac{1}{2(k+h^{\vee})}\sum_a t_\lambda^at_\lambda^a|\phi_\lambda\rangle 
\end{split}
\end{equation}
so that $|\phi_\lambda\rangle$ is a Virasoro primary of weight
\begin{equation}
    h = \frac{1}{2(k+h^{\vee})}\sum_a t_\lambda^at_\lambda^a \,.
\end{equation}
Notably, the converse is not true; not every Virasoro primary field is a WZW primary. For instance the current components $J^a(z)$ are Virasoro primaries, but they are not WZW primaries, which can be seen directly from the fact that the OPE \eqref{ec:currentalgebraOPE} does not take the form \eqref{ec:wzwprimaryope}.

\subsubsection{The Knizhnik-Zamoldchikov Equation}

The fact that WZW primary fields are also Virasoro primaries provides a set of additional constraints which allows us to translate the problem of computing correlation functions into a differential equation called the \textit{Knizhnik-Zamoldchikov equation}. To see how this works, we consider a WZW primary state $|\phi_{\lambda}\rangle$ and act on it with $L_{-1}$. From equation \eqref{ec:WZWstresstensor} we have
\begin{equation}
\begin{split}
    L_{-1}|\phi_{\lambda}\rangle 
    &= \frac{1}{k+h^{\vee}}\sum_a J_{-1}^a J^a_{0}|\phi_{\lambda}\rangle \\
    &=-\frac{1}{k+h^{\vee}}\sum_a J_{-1}^a t_{\lambda}^a|\phi_{\lambda}\rangle \,,
\end{split}
\end{equation}
so that the state
\begin{equation}
\label{ec:zerovector}
    |\chi\rangle=\left[L_{-1}+\frac{1}{k+h^{\vee}}\sum_a J_{-1}^a t_{\lambda}
    ^a\right]|\phi_{\lambda}\rangle=0
\end{equation}
is indeed, the zero vector. Now, in the same way than we did in \S \ref{sec:descendantfields} we may associate to $J_{-1}^a|\phi_{\lambda}\rangle$ a descendant field
\begin{equation}
  J_{-1}^a|\phi_{\lambda}\rangle = \phi^{(-1,a)}_{\lambda}(0)|0\rangle \quad \text{with}\quad \phi_{\lambda}^{(-1,a)}(w)=\frac{1}{2\pi i}\oint_w dz \frac{1}{z-w}J^a(z)\phi_{\lambda}(w) \,.
\end{equation}
If we insert it inside a correlation function of a set of WZW primaries, we get:
\begin{equation}
\begin{split}
    \langle \phi_{\lambda_1}(z_1)\dots \phi_\lambda^{(-1,a)}(w)\dots \phi_{\lambda_n}(z_n)\rangle 
    &=\frac{1}{2\pi i}\oint_{w}\frac{dz}{z-w}\langle J^a(z)\phi_{\lambda_1}(z)\dots \phi_{\lambda_n}(z)\rangle\\
    &=\frac{1}{2\pi i}\oint_{z_j\neq w}\frac{dz}{z-w}\sum_{j}\frac{t^a_\lambda}{z-z_j}\langle \phi_{\lambda_1}(z_1)\dots \phi_{\lambda_n}(z_n)\rangle\\
    &=\sum_{j}\frac{t^a_\lambda}{w-z_j}\langle \phi_{\lambda_1}(z_1)\dots \phi_{\lambda_n}(z_n)\rangle
\end{split}
\end{equation}
where in the first line we used the definition of the descendant field, and in the second we reversed the contour of integration in order to include every $z_j\neq w$. Hence, we obtain the expression
\begin{equation}
\begin{split}
    \langle \phi_{\lambda_1}(z_1)\dots \chi(w)\dots \phi_{\lambda_n}\rangle=\left[\partial_{w}+\frac{1}{k+h^{\vee}}\sum_{j}\frac{\sum_a t_\lambda^a\otimes t^a_{\lambda_j}}{w-z_j}\right]\langle \phi_{\lambda_1}(z_1)\dots \phi_{\lambda_n}(z_n)\rangle \,.
\end{split}
\end{equation}
But the correlator on the LHS vanishes because $\chi$ is the field associated with the zero vector (equation \eqref{ec:zerovector}), so we get the Knizhnik-Zamoldchikov differential equation \cite{Dashen,Knizhnik}:
\begin{equation}
\label{ec:kzequation}
    \left[\partial_{w}+\frac{1}{k+h^{\vee}}\sum_{j}\frac{\sum_a t_{\lambda}^a\otimes t^a_{\lambda_j}}{w-z_j}\right]\langle \phi_1(z_1)\dots \phi_n(z_n)\rangle=0\,.
\end{equation}
The solutions to this equation are the correlation functions of WZW primaries. As in the purely conformal case, the correlation functions of descendant fields can always be reduced to correlation functions of the corresponding primaries, so that solving this equation provides an exact solution for the theory. We observe how the existence of an additional infinite dimensional symmetry, which results in an extension of the chiral algebra, provides a set of additional constraints which lead to the exact solvability of the WZW model.

\section*{Acknowledgements}

This project started while completing Part III at the University of Cambridge, as a member of Hughes Hall College. I would like to thank Nick Dorey, Lewis Cole, Horacio Falomir, Pablo Pisani and Lucas Manso for the delightful discussions and suggestions. In particular, to H.F. and L.C. for carefully reading the manuscript and suggesting improvements. All of my work is supported by CONICET, through a PhD Scholarship.

\appendix

\section{Central Charge}
\label{sec:apendicecentral}

The central charge plays a very important role in the description of two dimensional conformal field theories, given its appearance in the commutation relations of the Virasoro algebra, which we used to structure our Hilbert space of states. Furhtermore, as we stated without proof, different values of the central charge can be identified with well known Lagrangian theories. For instance, we have that $c=1$ corresponds to the free boson and $c=1/2$ to the free fermion. Remarkably, one of the distinguished features of two dimensional conformal field theories, is the possibility of defining theories that do not admit a Lagrangian description, which in fact, are constructed based solely on symmetry principles. In this case, the allowed values of the central charge are constrained by the requirement of unitarity and positive norm states. Thus, we are lead to the following question:
\begin{center}
    \textit{Why is there a central charge?}
\end{center}
As a first approach to answering this question, we could refer to the theories with a known Lagrangian description, which indeed require the presence of a non vanishing central charge. We introduced this idea in \S \ref{sec:OPE} when stating the self OPE of the stress tensor, but we did not do any explicit calculation to show where the central charge term actually came from, we simply made reference to where those computations could be found in the literature. Along with its appearance, the physical interpretation of the central charge is discussed in several places \cite{Affleck, Blote, Cardy} (see for instance \cite{tong} for a very clear presentation) and it is usually related to a soft breaking of conformal symmetry due to the introduction of a macroscopic scale into the system when passing from the classical theory to the quantum theory. This, for example, can be related to a scale dependent regulation prescription, necessary to obtain a well defined QFT or to the introduction of scale dependent boundary conditions. In any case, the appearance of the central charge is strongly related to the Lagrangian description of the theory. So how can we explain the presence of the central charge in theories that do not admit a Lagrangian description?

To answer this question, we present, following \cite{Schottenloher}, a rather abstract mathematical argument, which has the virtue of being completely general. The underlying idea may be summarized as follows: Given a group of symmetries through a standard realization on some space we need to lift the action to the adequate geometrical objects we work with in our physical theory and sometimes, this action cannot be lifted. Mathematically, this is called an obstruction to the action of lifting. Obstructions often lead to the possibility of the realization not of the group of symmetries itself, but some extension of it by another group acting naturally on the geometrical objects defining the theory. This is precisely what happens in the process of quantization of the local conformal algebra in two dimensions, as we describe below. 

\subsection{Obstructions and the Virasoro Algebra}

The description of an obstruction in the context of representation theory requires going into some technicalities. We will try to make this section as friendly as possible including only the necessary concepts, but the use of some unfamiliar language may occur. However, before going into the details, we outline the idea we wish the reader to keep in mind when reading this section. We will talk about quantization, and therefore, we will be dealing with a Hilbert space $\mathcal{H}$. However, the Hilbert space \textit{per se} will not quite work because we want vectors that differ by a complex phase to be regarded as the same element (see for instance \cite{Weinberg} chap 1. for a detailed discussion). This requires us to quotient out all of the equivalent vectors leaving us with a well defined quantum phase space. At this point we are actually changing the mathematical objects we are dealing with, passing from a Hilbert space, to a quotient space. We will then consider the action of a symmetry group $G$ acting on the classical level, which translates into a unitary representation on the Hilbert space. The nuance occurs when we try to extend this notion to the quotient space, where, as we will discuss, we will find a mathematical obstruction. This obstruction, in the case of the local conformal algebra, will require us to replace the Witt algebra with the Virasoro algebra to obtain well defined unitary representations. Remarkably, from this perspective, the modification of the local symmetry algebra is not related to the introduction of an external scale breaking conformal invariance, but instead, to the requirement of dealing with well defined mathematical objects. 

The first step in quantizing a system is to define a Hilbert space $\mathcal{H}$ such that the quantum states are represented by vectors $|\psi\rangle \in \mathcal{H}$. There is however a very important constraint related to our probabilistic interpretation of the quantum theory, which is the fact that two vectors that differ by a complex phase are considered to be physically equivalent. This condition may be expressed mathematically by considering the projective space of one dimensional linear subspaces of $\mathcal{H}$, also known as \textit{rays}:
\begin{equation}
    \mathbb{P}(\mathcal{H}):=\left(\mathcal{H}\backslash \{0\}\right)/\sim
\end{equation}
where the equivalence relation $\sim$ is given by:
\begin{equation}
    [\psi] \sim [\psi']
\quad \Longleftrightarrow \quad |\psi\rangle = \lambda |\psi'\rangle \quad \text{for some } \lambda\in U(1)\,.
\end{equation}
Note that we are denoting by $[\psi]$ the equivalence classes in the quotient space, with $[\psi]=\gamma(|\psi\rangle)$ where $\gamma:\mathcal{H}\backslash\{0\}\to \mathbb{P}(\mathcal{H})$ is the canonical projection into the quotient. In particular, we have that any two vectors in $\mathcal{H}$ that differ by a complex phase belong to the same equivalence class in the quotient, and thus we take $\mathbb{P}(\mathcal{H})$ to be the quantum phase space. Now the physically measurable quantities are given by the transition probabilities between elements in $\mathbb{P}(\mathcal{H})$. Explicitly, if $[\psi]=\gamma(|\psi\rangle)$ and $[\psi']=\gamma(|\psi'\rangle)$, we have:
\begin{equation}
    P([\psi] \to [\psi'])=\frac{\langle \psi |\psi'\rangle}{\|\psi\|^2\|\psi'\|^2} \,.
\end{equation}
We define a symmetry transformation $T$ as a bijective mapping $T:\mathbb{P}(\mathcal{H})\to \mathbb{P}(\mathcal{H})$ that preserves the transition probability. We call these mappings \textit{projective transformations} and we denote by ${\rm Aut}(\mathbb{P})$ the set of all projective transformations. This set, together with composition, forms a group. The relationship between projective transformations defined over $\mathbb{P}(\mathcal{H})$ and the original Hilbert space $\mathcal{H}$ is given by Wigner's theorem \cite{Wigner}, which asserts that every projective transformation is \textit{induced} by either a unitary or anti-unitary operator acting on $\mathcal{H}$. To understand this last statement, let us introduce some definitions. Let $U(\mathcal{H})$ be the group of unitary operators over $\mathcal{H}$. For every $A\in U(\mathcal{H})$, we define a map $\hat{\gamma}(A):\mathbb{P}(\mathcal{H})\to \mathbb{P}(\mathcal{H})$ by:
\begin{equation}
    \hat{\gamma}(A)([\psi])=\gamma (A |\psi\rangle)
\end{equation}
for every $[\psi]=\gamma(|\psi\rangle)$. It can be shown that $\hat{\gamma}(A) \in {\rm Aut}(\mathbb{P})$, that is, $\hat{\gamma}(A)$ is bijective and preserves the transition probability. In terms of these definitions, Wigner's theorem states that for every projective transformation $T \in {\rm Aut}(\mathbb{P})$ there exist a unitary (or anti-unitary) operator $A\in U(\mathcal{H})$ such that 
\begin{equation}
    T= \hat{\gamma}(A) \,,
\end{equation}
which means that $T$ is induced by $A$. Moreover, we define
\begin{equation}
    U(\mathbb{P}):=\hat{\gamma}(U(\mathcal{H}))\subset {\rm Aut}(\mathbb{P})
\end{equation}
to be the group of \textit{unitary projective} transformations. Note that up to this point we have only considered single symmetry transformations, so our next step is to extend the analysis to a whole symmetry group $G$. The natural generalization, is to consider a \textit{unitary projective representation}; that is, a map:
\begin{equation}
    T:G\to U(\mathbb{P})\,,
\end{equation}
such that for every $g\in G$ we have that $T(g)\in U(\mathbb{P})$ is a unitary projective transformation, and furthermore, the group structure of $G$ is preserved (i.e. $T$ is a group homomorphism). The question is if Wigner's theorem can be generalized when we replace one symmetry transformation with a whole symmetry group $G$: Is every unitary projective representation induced by a unitary representation $R:G\to U(\mathcal{H})$, with $R$ a group homomorphism? Explicitly, given a unitary projective representation $T:G\to U(\mathbb{P})$, can we \textit{lift} it to a unitary representation $R:G\to U(\mathcal{H})$ such that $T= \hat{\gamma}\circ R$? The answer is \textit{no} in general. However, such a lift is possible if we weaken our assumptions, which leads us to the notion of a central extension. Informally, a central extension of a group $G$ is an extended group $E \supset G$ which has been augmented by another group $A$, such that the elements of $A \subset E$ commute with the elements of the original group $G\subset E$. Before going into formalities, let us think why would a central extension could come to our rescue. We started out with a Hilbert space $\mathcal{H}$ and we quotiented out all of the elements related by a complex phase, i.e, by an element of $U(1)$. \textit{Equivalently}, we may think of $\mathcal{H}$ as $\mathbb{P}(\mathcal{H})$ augmented by the elements we have quotiented out, i.e, by $U(1)$. When passing from the Hilbert space (and its quotient) to unitary transformations defined over these spaces, we require some machinery that is capable of realizing that at the level of the Hilbert space, a quotient has been taken. This is exactly the role played by central extensions, which deals with groups from an abstract perspective. Let us see how can we express this idea mathematically. An extension of $G$ by a group $A$ is given by the \textit{exact sequence} of group homomorphisms
\begin{equation}
\label{ec:exactsequence}
\begin{tikzcd}
1 \arrow[r] & A \arrow[r, "\iota"] & E \arrow[r, "\pi"] & G \arrow[r] & 1
\end{tikzcd}
\end{equation}
where exactness means that the kernel of each map in the sequence equals the image of the previous map. Let us bring this down to earth. The fact that the kernel of $\iota$ is the image of the trivial map $1 \to A$ is a fancy way of saying that $\iota$ is injective and thus, that we may think of $A$ as a subset of $E$. Similarly, the fact that the image of $\pi$ is the kernel of the trivial map $G\to 1$ is a fancy way of saying that $\pi$ is surjective and thus, we may think of $G$ as a subset of $E$. In this case, we say that $E$ is an extension of $G$ by $A$. Furthermore, the extension is \textit{central} if $\iota(A) \subset E$ is a subset of the center of $E$, i.e, if every element of $A$ thought as a subset of $E$ commutes with every element of $E$. With these definitions, we are now in conditions to state the generalization of Wigner's theorem to a whole symmetry group $G$: Given a group $G$ and unitary projective representation $T:G\to U(\mathbb{P})$, there is a central extension $E$ of the universal covering group $\tilde{G}$ of $G$ by $U(1)$, and a unitary representation $R:E \to U(\mathcal{H})$ such that $\hat{\gamma}\circ R=T\circ \pi$, where $\pi$ is the map defining the central extension of $\tilde{G}$ by $U(1)$ as in equation \eqref{ec:exactsequence}. In other words, \textit{every unitary projective representation is induced by a unitary representation of a central extension of the universal covering group} $\tilde{G}$ \textit{by} $U(1)$. 

To make contact with something that we already know, let us consider an example of this procedure, which appears when studying rotations in three dimensional space. The group of rotations in three dimensions is $SO(3)$ and it's universal covering group is $SU(2)$. Then the theorem states that unitary projective representations of $SO(3)$ are induced by unitary representation of a central extension of $SU(2)$ by $U(1)$.  Notably, every central extension of $SU(2)$ by $U(1)$ is trivial, in the sense that is given by the exact sequence:
\begin{equation}
\label{ec:exactsequence}
\begin{tikzcd}
1 \arrow[r] & U(1) \arrow[r, "\iota"] & SU(2)\times U(1) \arrow[r,"\pi_1"] & SU(2) \arrow[r] & 1 \,,
\end{tikzcd}
\end{equation}
where $\iota(a)=(\mathrm{Id},a)$ and $\pi_1(a,b)=a$. Thus, a unitary projective representation $T:SO(3)\to U(\mathbb{P})$ can be lifted to a unitary representation $S:SU(2)\times U(1) \to U(\mathcal{H})$ which factorizes into a unitary representation $R:SU(2)\to U(\mathcal{H})$ with $S=R\circ \pi_1$. This is the reason why, when considering three dimensional rotations in the quantum theory, we take unitary representations defined over $SU(2)$ instead of $SO(3)$ as in the classical theory. Exactly the same thing happens in the analysis of Lorentz transformations, where we replace the symmetry group $SO(1,3)$ by the (trivial) central extension of the universal covering group $SL(2,\mathbb{C})$. 

When studying conformal field theories in two dimensions, we instead use Lie algebras to describe symmetries, given the importance of local conformal transformations (c.f \S \ref{sec:mathematicalaspectsd2}), which are better described from the Lie algebra persepective. Thus, we are interested in understanding how the above considerations for groups can be repeated in terms of Lie algebras. 

Let us start by defining the analogous concepts of unitary and (unitary) projective representations for Lie algebras. Let $\frak{u}(\mathcal{H})$ be the Lie algebra of self-adjoint operators on $\mathcal{H}$ and let $\frak{u}(\mathbb{P})$ be the Lie algebra of $U(\mathbb{P})$. Then a unitary projective representation of a Lie algebra $\frak{g}$ is a Lie algebra homomorphism $t:\frak{g}\to \frak{u}(\mathbb{P})$ and similarly, a unitary representation is simply a Lie algebra homomorphism $r:\frak{g}\to \frak{u}(\mathcal{H})$. In the same way than with groups, symmetries will be given by unitary projective representations of the symmetry algebra, and thus the question is under what circumstances these may be lifted to unitary representations. To answer this question, we introduce some definitions. A central extension $\frak{b}$ of a Lie algebra $\frak{c}$ by a Lie algebra $\frak{a}$ is an exact sequence of Lie algebras
\begin{equation}
\label{ec:centralextensionalgebras}
\begin{tikzcd}
0 \arrow[r] & \frak{a} \arrow[r, "\alpha"] & \frak{b} \arrow[r, "\beta"] & \frak{c} \arrow[r] & 0
\end{tikzcd}
\end{equation}
with $\frak{a}$ abelian and $[X,Y]=0$ for every $X \in \frak{a}$ and $Y\in \frak{b}$ where we identify $\frak{a}$ with the corresponding subalgebra of $\frak{b}$. We may relate central extensions of Lie algebras with central extensions of Lie groups as follows. Let 
\begin{equation}
\begin{tikzcd}
1 \arrow[r] & A \arrow[r, "I"] & E \arrow[r, "R"] & G \arrow[r] & 1
\end{tikzcd}
\end{equation}
be a central extension of Lie groups with differentiable\footnote{Differentiability is required as to preserve the differentiable manifold structure of the Lie group} homomorphisms $I$ and $R$. Then for $\dot{I}={\rm Lie}\, I$ and $\dot{R}={\rm Lie}\, R$, the Lie algebra homomorphisms induced by $I$ and $R$ respectively, we have that the sequence 
\begin{equation}
\begin{tikzcd}
0 \arrow[r] & {\rm Lie}\, A \arrow[r, "\dot{I}"] & {\rm Lie}\, E \arrow[r, "\dot{R}"] & {\rm Lie}\, G \arrow[r] & 0
\end{tikzcd}
\end{equation}
is a central extension of Lie algebras. Of particular interest for our discussion, are the central extensions by $U(1)$: The central extension of groups
\begin{equation}
\begin{tikzcd}
1 \arrow[r] & U(1) \arrow[r] & U(\mathcal{H}) \arrow[r, "\hat{\gamma}"] & U(\mathbb{P})\arrow[r] & 1
\end{tikzcd}
\end{equation}
induces a central extension of Lie algebras:
\begin{equation}
\label{ec:u1centralextension}
\begin{tikzcd}
0 \arrow[r] & \frak{u}(1)\cong \mathbb{R} \arrow[r] & \frak{u}(\mathcal{H}) \arrow[r,"\tilde{\gamma}"] & \frak{u}(\mathbb{P})\arrow[r] & 0\,.
\end{tikzcd}
\end{equation}
As we have seen in section \ref{sec:mathematicalaspectsd2} the local conformal algebra in both Euclidean and Minkowski space contains two copies of the Witt algebra, which we will denote by $\mathrm{W}$. Thus, given a unitary projective representation $t:\mathrm{W}\to \frak{u}(\mathbb{P})$, with $t$ a Lie algebra homomorphism, we ask under what circumstances we may lift $t$ to a unitary representation $r:\tilde{W}\to \frak{u}(\mathcal{H})$ for some Lie algebra $\tilde{W}$. In the case of groups, we saw that unitary projective representations could always be lifted to unitary representations of a $U(1)$ central extension of the universal covering of the original group. For Lie algebras we have a completely analogous result; every unitary projective representation of a Lie algebra $\frak{g}$ can be lifted to a unitary representation of a central extension\footnote{Note that for Lie algebras the lifting is done with respect to the central extension of the original algebra directly.} of $\frak{g}$ by $\frak{u}(1)$. Hence, if we want to lift projective representations of the Witt algebra we should consider the $\frak{u}(1)$ central extensions of $\mathrm{W}$. Notably, the Witt algebra has essentially a unique non trivial central extension by $\frak{u}(1)$ and it's given by the Virasoro algebra \cite{Gelfand}. Thus, every unitary projective representation $t:\mathrm{W}\to \frak{u}(\mathbb{P})$ lifts to a unitary representation of the Virasoro algebra $r:\mathrm{Vir}\to \frak{u}(\mathcal{H})$. Explicitly, we have the following commutative diagram
\begin{equation}
\begin{tikzcd}
0 \arrow[r] & \frak{u}(1) \arrow[r] \arrow[d] & \mathrm{Vir} \arrow[r, "\pi"] \arrow[d, "r"]      & \mathrm{W} \arrow[r] \arrow[d, "t"] & 0 \\
0 \arrow[r] & \frak{u}(1) \arrow[r] & \frak{u}(\mathcal{H}) \arrow[r, "\tilde{\gamma}"] & \frak{u}(\mathbb{P}) \arrow[r]      & 0
\end{tikzcd}
\end{equation}
where $\pi$ is the projection of $\mathrm{Vir}$ into $\mathrm{W}$ and $\tilde{\gamma}$ is the induced Lie algebra homomorphism defined in equation \eqref{ec:u1centralextension}. Commutativity of the diagram implies $t\circ \pi = \tilde{\gamma}\circ r$ and thus, we have the lifting of the projective representation. In particular, the Virasoro algebra is a $\frak{u}(1)\cong \mathbb{R}$ central extension of the Witt algebra and thus we may write:
\begin{equation}
    \mathrm{Vir}=\mathrm{W}\oplus \mathbb{R}\,.
\end{equation}
In other words, the Virasoro algebra is the Witt algebra augmented with an additional generator that commutes with every generator of $\mathrm{W}$ and that it can be characterized by a real number $c$; the central charge. The defining commutation relations of $\mathrm{Vir}$ are:
\begin{align*}
    [L_n,L_m]
    &=(n-m)L_{n+m}+\frac{c}{12}n(n^2-1)\delta_{n+m,0}\\
    [L_n,c\,]&=0\,.
\end{align*}
We have argued that the appearance of the Virasoro algebra as the adequate structure to describe local conformal symmetry in two dimensional quantum field theories may be understood from the requirement of obtaining well defined unitary representations induced by projective ones. This perspective, which certainly differs from the one presented in \S \ref{sec:OPE}, can be applied to systems with conformal symmetry without the need of making reference to a Lagrangian description, but of course, it can be equally applied to those systems as well!  

\section{State Operator Correspondence and the OPE}
\label{sec:apendiceOPE}

Let us argue why in conformal field theories, the OPE provides a convergent series at finite point separation. Let $\mathcal{O}_i$ and $\mathcal{O}_j$ be a pair of local operators and let us consider the correlation function of it's time ordered product $\langle \mathcal{O}_i(z,\bar{z})\mathcal{O}_j(0,0)\rangle$. In the context of radial quantization, we may interpret this correlation function as the vacuum expectation value of the radially ordered product of fields:
\begin{equation}
\label{ec:opeexact1}
    \langle \mathcal{O}_i(z,\bar{z})\mathcal{O}_j(0,0)\rangle = \langle 0|\mathcal{R}\left[\mathcal{O}_i(z,\bar{z})\mathcal{O}_j(0,0)\right]|0\rangle\,. 
\end{equation}
Now the state $\mathcal{O}_i(z,\bar{z})\mathcal{O}_j(0,0)|0\rangle = |\Psi\rangle$ will be a certain vector in a Hilbert space defined over $z=0$. Thus, we may expand it as a linear combination of basis vectors of the Hilbert space:
\begin{equation}
\label{ec:opeexact2}
    |\Psi\rangle = \sum_{k}C_{ij}^k(z,\bar{z})|\mathcal{O}_k\rangle\,.
\end{equation}
Because of the operator state correspondence we have that for every basis state $|\mathcal{O}_k\rangle$ there exist a unique local operator $\mathcal{O}_k(z,\bar{z})$ corresponding to it. Hence, from equation \eqref{ec:opeexact1} we have:
\begin{equation}
\begin{split}
        \langle \mathcal{O}_i(z,\bar{z})\mathcal{O}_j(0,0)\rangle 
        &=\langle 0|\mathcal{O}_i(z,\bar{z})\mathcal{O}_j(0,0)|0\rangle\\
        &= \langle 0|\sum_k C_{ij}^k(z,\bar{z})|\mathcal{O}_k\rangle\\
        &=\langle 0|\sum_k C_{ij}^k(z,\bar{z})\mathcal{O}_k(0,0)|0\rangle\\
        &=\sum_k C_{ij}^k(z,\bar{z})\langle\mathcal{O}_k(0,0)\rangle\,, \\
       \end{split}    
\end{equation}
where in the second line we replaced expression \eqref{ec:opeexact2}, and in the third line we used the state operator correspondence. Now the crucial point is that the sum in the second line is done over a Hilbert space, which by definition is \textit{complete}. This, in turn, implies that the sum converges\footnote{We recall that a normed space is complete if and only if every Cauchy sequence converges. A sequence $\{a_n\}$ is a Cauchy sequence if for every $\epsilon>0$ there exists an $n_0 \in \mathbb{N}$ such that $|a_{n+1}-a_n|<\epsilon$
for every $n\geq n_0$. In general, this condition does not imply that the sequence converges, except in the case in which the space is complete. In particular, given that the OPE is an asymptotic expansion, we may define a Cauchy sequence by considering the partial sums $a_n=\sum_{k}^nC_{ij}^k(z,\bar{z})|\mathcal{O}_k\rangle$. These will satisfy the Cauchy sequence condition, and completeness of the Hilbert space will then guarantee the convergence of the OPE.}. Of course, convergence of the sum in the last line follows immediately, which shows that the OPE (as defined in \eqref{ec:OPE123}) is exact. Notably, exactness of the OPE strongly relies on the state operator correspondence, which in turn its a consequence of the conformal mapping from the cylinder to the plane.

\printbibliography[title=\textsc{References}]

\end{document}